\documentclass[12pt]{iopart}
\usepackage{graphicx}

\begin{document}

\title
[]
{Analysis of EEG signal by Flicker Noise Spectroscopy: 
Identification of right/left hand movement imagination}

\author
{A. Broniec$^1 $ }

\address
{$^1$AGH University of Science and Technology, Faculty of Electrical Engineering, Automatics, Computer Science and Biomedical Engineering, 
al. A. Mickiewicza 30, Krak\'ow, Poland}

\ead{abroniec@agh.edu.pl}


\submitto{}

\maketitle

\begin{abstract}

Flicker Noise Spectroscopy (FNS) has been used for the analysis of
electroencephalography (EEG) signal related to the movement imagination.
The analysis of sensorimotor rhythms in time-frequency maps reveals the
event-related desynchronization (ERD) and the post-movement
event-related synchronization (ERS), observed mainly in the
contralateral hemisphere to the hand moved for the motor imagery. The
signal has been parameterized in accordance with FNS method. The
significant changes of the FNS parameters, at the time when the subject
imagines the movement, have been observed. The analysis of these
parameters allows to distinguish between imagination of right and left
hands movement. Our study shows that the flicker-noise spectroscopy can
be an alternative method of analyzing EEG signal related to the
imagination of movement in terms of a potential application in the
brain-computer interface (BCI). 

\end{abstract}

\section{Introduction}
\label{sec:1}

Neuropsychological studies on processes occuring in the brain during the
Motor Imagery (MI) show that similar parts of the brain are involved in
the movement imagination as well as its real performance
~\cite{Deecke1991,Deecke1996a,Deecke1996b,Kus2006}. The main difference
between the movement execution and its imaginary is that in the latter
case, the movement performance is blocked at some level of the
corticospinal information transfer
~\cite{Decety1988,Decety1990,Decety1993,Jeannerod1994,Jeannerod1995}.
This phenomenon is observed in the sport psychology, where many
examinations show that mental exercises have a positive effect on the
later movement execution ~\cite{Feltz1983,Denis1985,Rodgers1991}.
Similarity between the movement imagination and the real movement has
been also confirmed by the research, in which healthy patients and
patients with motor disabilities have been subjected to the
neuropsychological observation~\cite{Dominey1995, Neuper2003}. This fact
causes that the movement imagination 
plays an important role as a control signal in the brain-computer
interface (BCI)~\cite{Wolpaw2002}, which are dedicated to patients who
partly or entirely lost the voluntary muscle contraction such as in the
'locked-in' state~\cite{McFarland2000,Pineda2000,Kus2012}. 
The natural way of description both, the execution and the imagination
of movement is the event-related desynchronization (ERD) and the
event-related synchronization
(ERS)~\cite{Pfurtscheller1977,Pfurtscheller1979,Pfurtsheller1994,
HandbookEEG}. The spatiotemporal maps of ERD/ERS inform about the power
decrease/increase in the brain activity in mu (8$-$13~Hz) and beta
frequency range (15$-$30~Hz), averaged over trials with respect to the
power in a reference time interval~\cite{Durka2004,Zygierewicz2005}. In
case of experiment with hand movement, ERD appears in both mu and beta
bands before movement (imagination) while ERS appears usually in the
beta band as the post-movement beta synchronization ($\beta-$rebound).
Since human somatotopic organization indicates that human limbs are
controlled by contralateral brain hemispheres
~\cite{Stancak1996b,Stancak1996c,Ginter2005}, we expect that the most
important changes in the brain activity during right hand movement
imagination occur mainly at electrode C3 which 
lies over the left hemisphere of the motor cortex whereas during left
hand movement imagination at electrode laying over the right brain
hemisphere i.e.~C4.

Flicker Noise Spectroscopy (FNS) is a time series analysis method that
introduces parameters characterizing the components of stochastic
signals in different frequency ranges. So far the method has been
applied to the parameterization of images produced by the atomic force
microscopy (AFM)~\cite{Timashev2006a}, analysis of geological signals
measured in seismic areas (determination of earthquake
precursors)~\cite{Descherevsky2003,Telesca2004}, determination of
electric breakdowns precursors in thin porous silicon
films~\cite{Parkhutik2003}, analysis of electric potential fluctuations
in electromembrane systems~\cite{Timashev2003}. FNS method has been also
successfully applied to some problems in medical data processing. It is
worth mentioning that the FNS method has been used to analyse the effect
of different types of medical treatment on the dynamics of index finger
tremor in Parkinsonian patients~\cite{Yulmetyev2006}. In
Ref.~\cite{Timashev2009} Timashev et al. have used FNS for the
identification of the photosensitive epilepsy. Their results suggest
that FNS is a promising method of early diagnosis, not only for the
photosensitive epilepsy but also for other neurodegenerative diseases
such as Parkinson's, Alzheimer's, Huntington's, amyotrophic lateral
sclerosis and schizophrenia. These suggestions have been confirmed in
Ref.~\cite{Timashev2012}, where it has been found that the FNS
parameterization of EEG signal may be used for the diagnosis of
schizophrenia at the early stages of its development.

Motivated by the successful applications of the flicker-noise
spectroscopy in the wide range of medical diagnoses, in the present
paper we use this method to find episodes related to the imagination of
the right and left hand movement. We have found that the changes in the
FNS parameters allows to determine the moment at which the movement
imagination occurs. It is also possible to differentiate which hand
participates in the task. This classification is crucial to BCI
applications.

The paper is organized as follows: in Sec.~\ref{sec:2}, we provide the
fundamentals of FNS and present the parameterization algorithm. In
Sec.~\ref{sec:3}, we present the experimental paradigm and methods of
EEG data acquisition. The results and discussion are presented in
Sec.~\ref{sec:4} while Sec.~\ref{sec:5} contains the conclusions.

\section{The flicker-noise spectroscopy methodology}
\label{sec:2}

The basic idea of FNS method is the assumption that the main
information about the system under study is provided by specific
"resonant" and "chaotic" components including sequences of different
types of irregularities such as spikes, jumps and discontinuities in
their derivatives of different orders~\cite{Timashev2001}. In FNS method
the analysed signal is separated into following components:
low-frequency corresponding to the system specific "resonances" and two
"chaotic" components having the source in irregularities of the signal,
appropriately spike and jumps.

The signal under consideration $V(t)$ (in our case $V(t)$ corresponds
to the EEG signal) can be written in the form
\begin{equation}
V(t)=V_r(t)+V_{cS}(t)+V_{cR}(t), \label{eq:1}
\end{equation}
where $V_r(t)$ is the low-frequency signal formed by resonant
component, $V_{cS}(t)$ is the chaotic component formed by spikes
(mostly, the highest-frequency band based on the concept of the Dirac
$\delta-$function) and $V_{cR} (t)$ is the chaotic component formed by
jumps (mostly, the intermediate-frequency band, "jumps-like" based on
the concept of the Heaviside  $\theta-$functions). According to FNS, the
main tool to extract and analyse the information contained in the signal
is the power spectrum (the cosine transform) of the autocorrelation
function defined as
\begin{equation}
S(f)=\int_{-T/2}^{T/2}\psi(\tau)\cos(2\pi f\tau)d\tau, \label{eq:2}
\end{equation}
where $\tau$ is the time lag $0\leq\tau\leq T/2$ and $\psi(\tau)$ is
the autocorrelation function which can be expressed in the form
\begin{equation}
\psi(\tau)=\frac{1}{T-\tau}\int_{0}^{T-\tau}V(t)V(t+\tau) dt.\label{eq:3}
\end{equation}
To extract the additional information contained in $\psi(\tau)$
($\left<V(t)\right>=0$ is assumed), the difference moments (Kolmogorov
transient structure function) of the second order $\phi^2(\tau)$ is
required
\begin{equation}
\phi^2(\tau)=\frac{1}{T-\tau}\int_{0}^{T-\tau} {\left[V(t)-V(t+\tau)\right]}^2 dt.\label{eq:5}
\end{equation}
Note that the functions $S(f)$ and $\phi^2(\tau)$ contain information
averaged over time interval~$T$. In real experiment $T$ can be a
subinterval of $T_{tot}$, where $T_{tot}$ corresponds to the whole
duration of the experiment. It is assumed that process in the $T$
interval is stationary. Thus, the time of the experiment $T_{tot}$ is
examined with the sliding window technique. The assumption that in the
time interval $T$ the process is stationary, leads to the following form
for the difference moments function
\begin{equation}
\phi^2(\tau)=2\left[\psi(0)-\psi(\tau) \right]\!.\label{eq:6}
\end{equation}

\subsection{Signal parameterization}

FNS method allows to determine several parameters which describe the
dynamics/characteristic of the system. All these parameters can be
extracted from chaotic components of the functions $S(f)$ and
$\phi^2(\tau)$ using appropriate interpolation formulas presented
underneath. The formula for the chaotic component of the difference
moments $\phi^2_c(\tau)$ is expressed as ~\cite{Yulmetyev2006}
\begin{equation}
\phi^2_c(\tau)\approx 2\sigma^2{\left[1-\Gamma^{-1}(H_1,\frac{\tau}{T_1})\right]}^2\!,\label{eq:7}
\end{equation}
where $\Gamma(s)=\Gamma(s,0)$, $\Gamma(s,x)=\int_{x}^{\infty}
\exp(-t)t^{s-1}dt$ are the complete and incomplete gamma functions
($x\geq0$ and $s>0$), $\sigma$ is the standard deviation of the measured
variable, $H_1$ is the Hurst constant and $T_1$ is the correlation time.
The interpolation function for chaotic power spectrum component $S_c
(f)$ is separated into two independent parts related to spikes
$S_{cS}(f)$ and jumps $S_{cR}(f)$
\begin{equation}
S_{cS}(f)=\frac{S_{cS}(0)}{1+{\left(2\pi fT_0 \right)}^{n_0}},\label{eq:8}
\end{equation}
\begin{equation}
S_{cR}(f)=\frac{S_{cR}(0)}{1+{\left(2\pi fT_1 \right)}^{2H_1+1}},\label{eq:9}
\end{equation}
where $S_{cS}(0)$, $n_0$, $T_0$ are the parameters and $S_{cR}(0)$ is expressed as 
\begin{equation}
S_{cR}(0)=4\sigma^2 T_1 H_1 \left\{1- \frac{1}{2H_1 \Gamma^2(H_1)} \int_0^{\infty}\Gamma^2(H_1,\xi)d\xi \right\}\!.\label{eq:10}
\end{equation}
Parameters introduced above have a determinate physical interpretation
and can be considered as parameters characterizing the signal $V(t)$.
Parameter $\sigma$ is the standard deviation of the signal, $H_1$ is the
Hurst constant, which describes the rate at which the dynamic variable
"forgets" its values on the time intervals that are less than the
correlation time $T_1$. Time $T_1$ determines the characteristic time
interval, during which the values of measured signal $V(t)$ stop
correlating. Parameter $S_{cS}(0)$ characterizes the boundary value of
$S_{cS}(f)$ in the low frequencies bands, $S_{cR}(0)$ characterizes the
boundary value of $S_{cR}(f)$ in the low frequencies bands, whereas
$n_0$ describes the degree of the correlation loss in the frequency
domain, when the frequency approach to the value $1/T_0$. More details
concerning FNS method can be found in
Refs.~\cite{Timashev2001,Timashev2003,Timashev2006a,Timashev2006b,
Timashev2007,Timashev2008,Timashev2009,Timashev2010a,Timashev2010b,
Timashev2012}.

In case of nonstationary processes, it is suggested to check the
dynamic of the parameters changes in the consecutive time windows
$[t_k,t_k+T]$ (where $k~=~0,1,2,3,\dots$ and $t_k=k\Delta T$), shifted
within the time limit of the total duration of the
experiment~\cite{Timashev2008}. It is assumed that in each window
$[t_k,t_k+T]$ the signal is stationary. This procedure is analogous to
the method of the sliding window applied in the classical technique of
the signal analysis.

\section{Methods}
\label{sec:3}
\subsection{Subjects and Data Acquisition}

\begin{figure}[ht]
\begin{center}
\includegraphics[scale=.4]{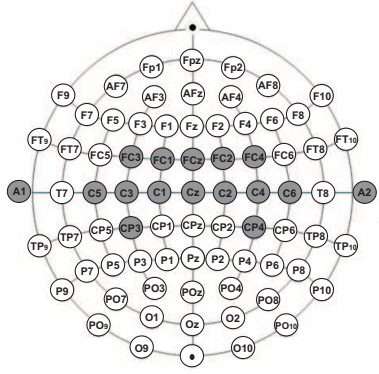}
\caption{Extended international $10-20$ system of the electrodes
placement. The electrodes used for the data acquisition are marked with
gray circle. The ear's electrodes $A_1$ and $A_2$ are the referencing
ones.}
\label{fig1}
\end{center}
\end{figure}

Three volunteers (two females and one male) between the ages of 24 and
35 participated in this study. Two of them are right handed and one
patient is bimanual. One of them suffers from spinal muscular atrophy
(SMA). All subjects gave informed consent. 
Each subject was seated in a comfortable armchair located about 1.5~m
in front of a computer screen. Subjects were requested to relax the
muscle and suppress eye blinking to avoid EMG and EOG activity
artefacts. The trials with evident artefacts were excluded and only
artifact-free EEG segments were used for the further analysis. Unipolar
EEG-channels were recorded from 14 gold disk electrodes placed over the
left and right hemisphere over the cortical hand area according to the
international extended $10-20$ system. Disk electrodes enable to keep
the resistance between electrodes in the range $0.1-3.5$~k$\Omega$. The
configuration of electrodes used for the data acquisition is shown in
Fig.~\ref{fig1}. All 14 channels were referenced to the right or left
ear's lobe signals and ground from the forehead. Signals from all the
channels were amplified with the biomedical signal amplifier g.USBamp
(USB Biosignal Amplifier g.tec Guger Technologies). EEG was band-pass
filtered between 0.5 and 100~Hz and recorded 
with a sample frequency of 1200~Hz.

\subsection{Experimental paradigm}
\begin{figure}[htb]
\begin{center}
\includegraphics[scale=.65]{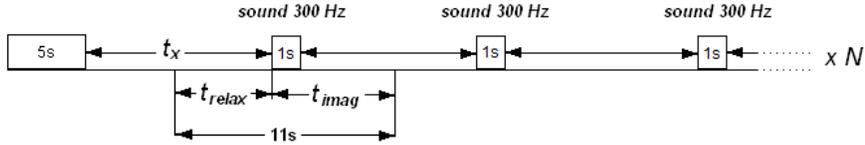}
\caption{Schematic diagram of the experimental paradigm. $t_x$ is the
time between the consecutive stimulus initiating the performance of the
task. Its value varies randomly between 10 and 15 seconds. $t_{relax}$
is a relaxation time preceding the execution of the task, $t_{imag}$ is
a time in which the imagination of the movement and the return to the
relaxation state occur. The time interval taken for the further analysis
lasts 11 seconds.}
\label{fig2}
\end{center}
\end{figure}

Since the motor imagery is not a routine natural behavior in a daily
life, usually the mental training with feedback is required before
subjects can perform a vivid imagination of the movement
~\cite{Neuper2009}. The schematic diagram of the experimental paradigm
is shown in Fig.~\ref{fig2}.
The experiment consists of 30 repetitions of right or left hand
movement imagery trials. The duration time of the single trial takes 11
seconds and consists of three periods. First period, from 0 to 5 s, is
the relaxation time used as the referential time needed for calculating
time-frequency maps of ERD/ERS changes. In the second period, from 5 to
6~s, the sound signal indicates the performance of the hand movement
imagination. The third one, from 6 to 11~s, is the time for the
execution of the task and return to the relaxation state. The duration
time between the consecutive stimuli varies randomly but is not shorter
than 8 seconds. This condition guarantees that the referential time is
not disturbed by the expectation of the stimulus. In the experiment the
duration time $t_x$ between each stimulus  is varied between 10 and
15~s. 
Each subject participated in at least two sessions. Each session
consists of $4-6$ runs with 30 trials $-$ half of them is for the right
and half for the left hand imagination. The number of runs depends on
the tiredness of the subject.

\section{Results and discussion}
\label{sec:4}

The ERD/ERS maps are the classical way of brain activity presentation
in the experiment in which the sensorimotor cortex behaviour is
investigated. In this section, we present the maps of ERD/ERS related to
the hand movement imagination in the time-frequency plane. The ERD/ERS
maps, calculated by the Continuous Wavelet Transform, are treated as the
reference point for the FNS analysis. Then the results of the FNS
parameterization methodology are presented and finally, the changes of
the FNS parameters as a function of time (during imagination of hand
movement) are shown.
\begin{figure}[ht]
\begin{center}
\includegraphics[scale=.17]{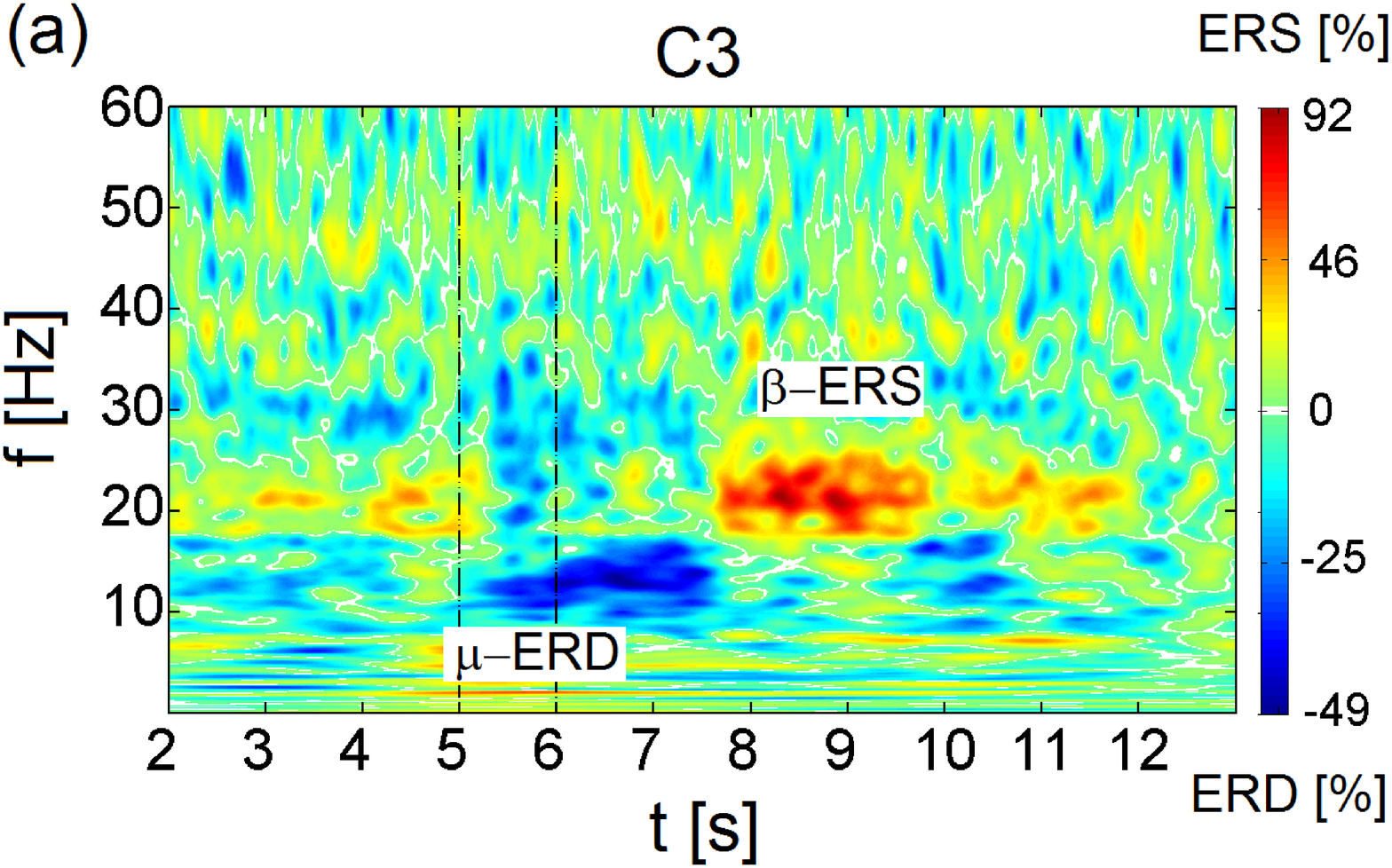}
\includegraphics[scale=.17]{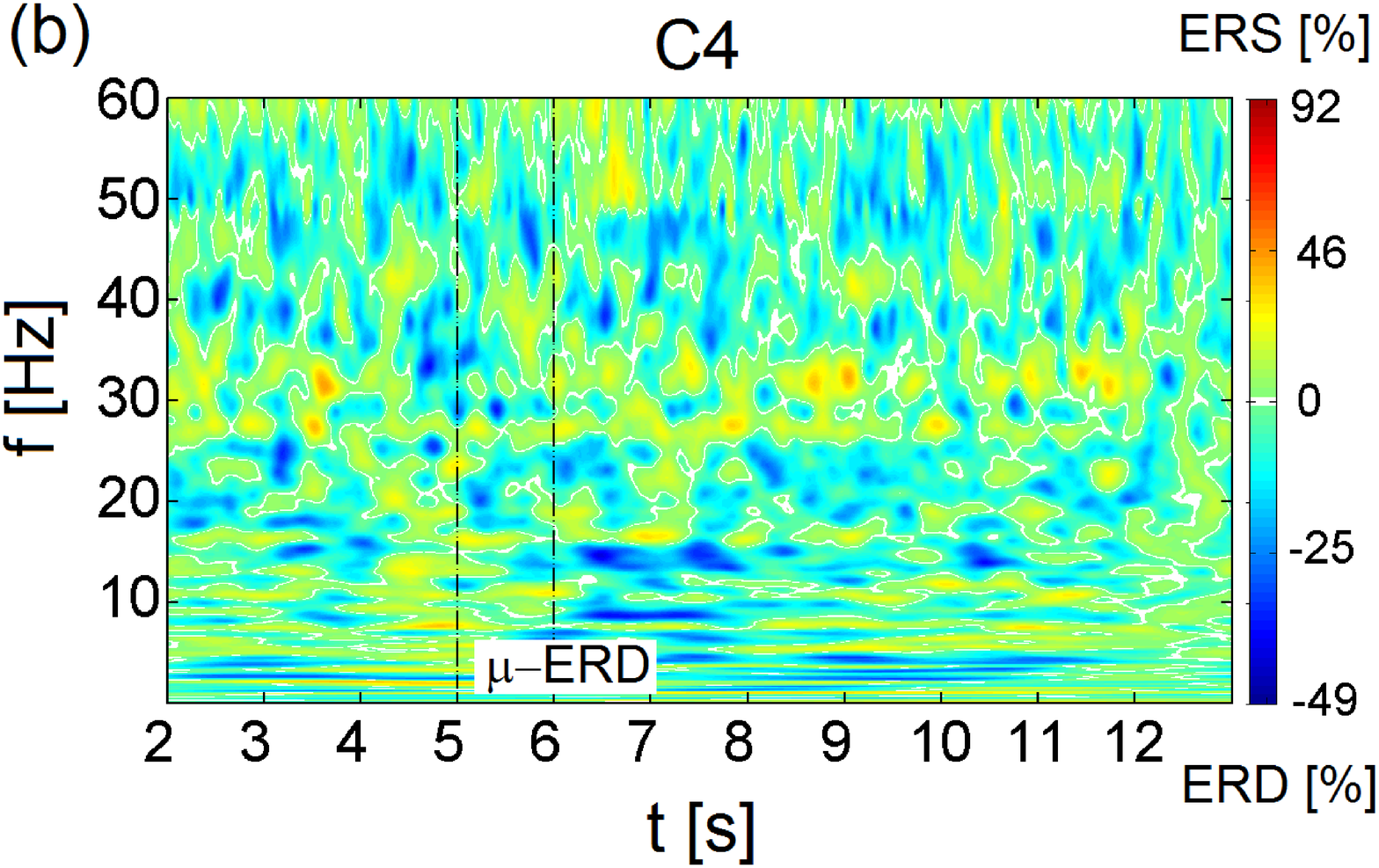}
\includegraphics[scale=.17]{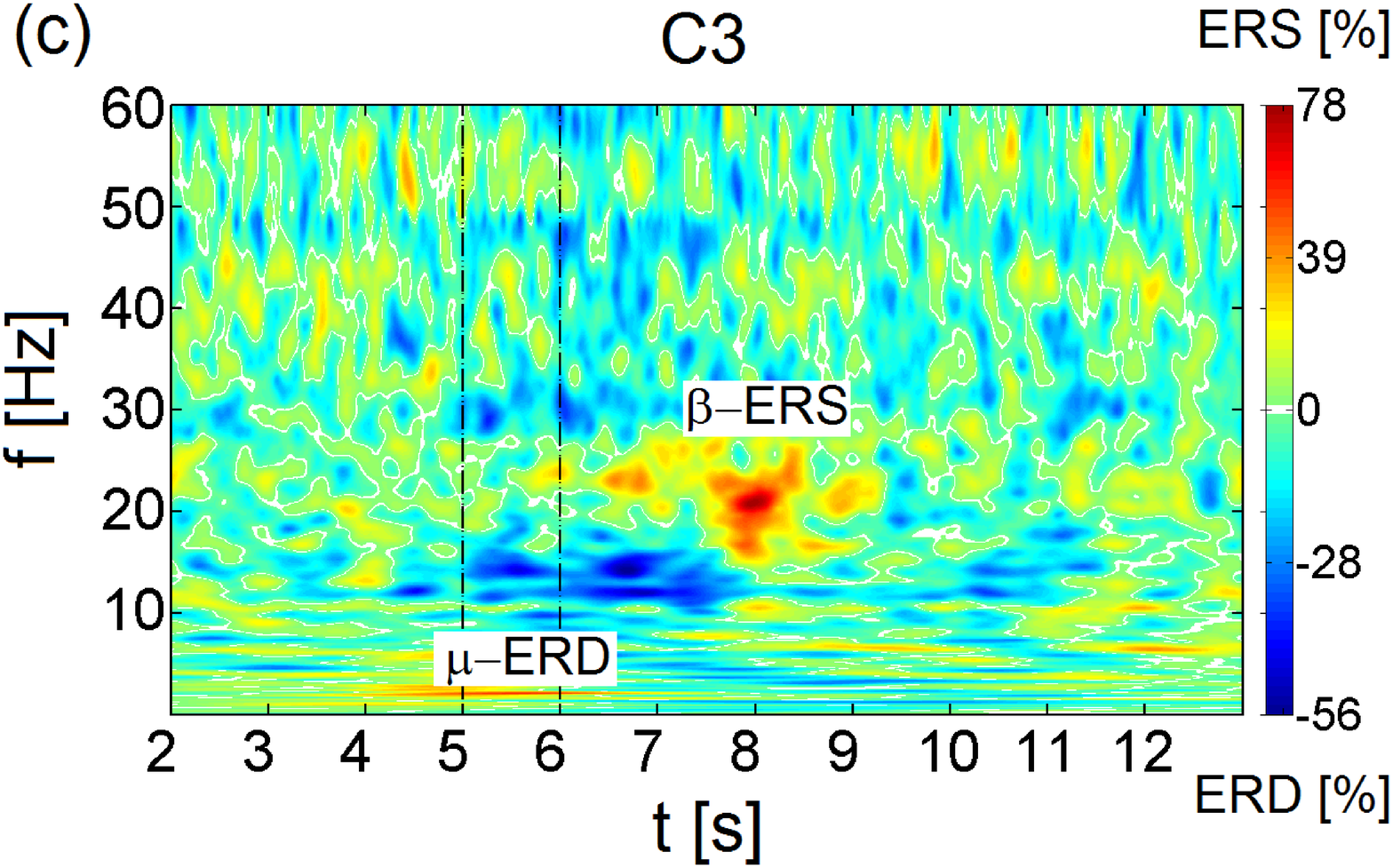}
\includegraphics[scale=.17]{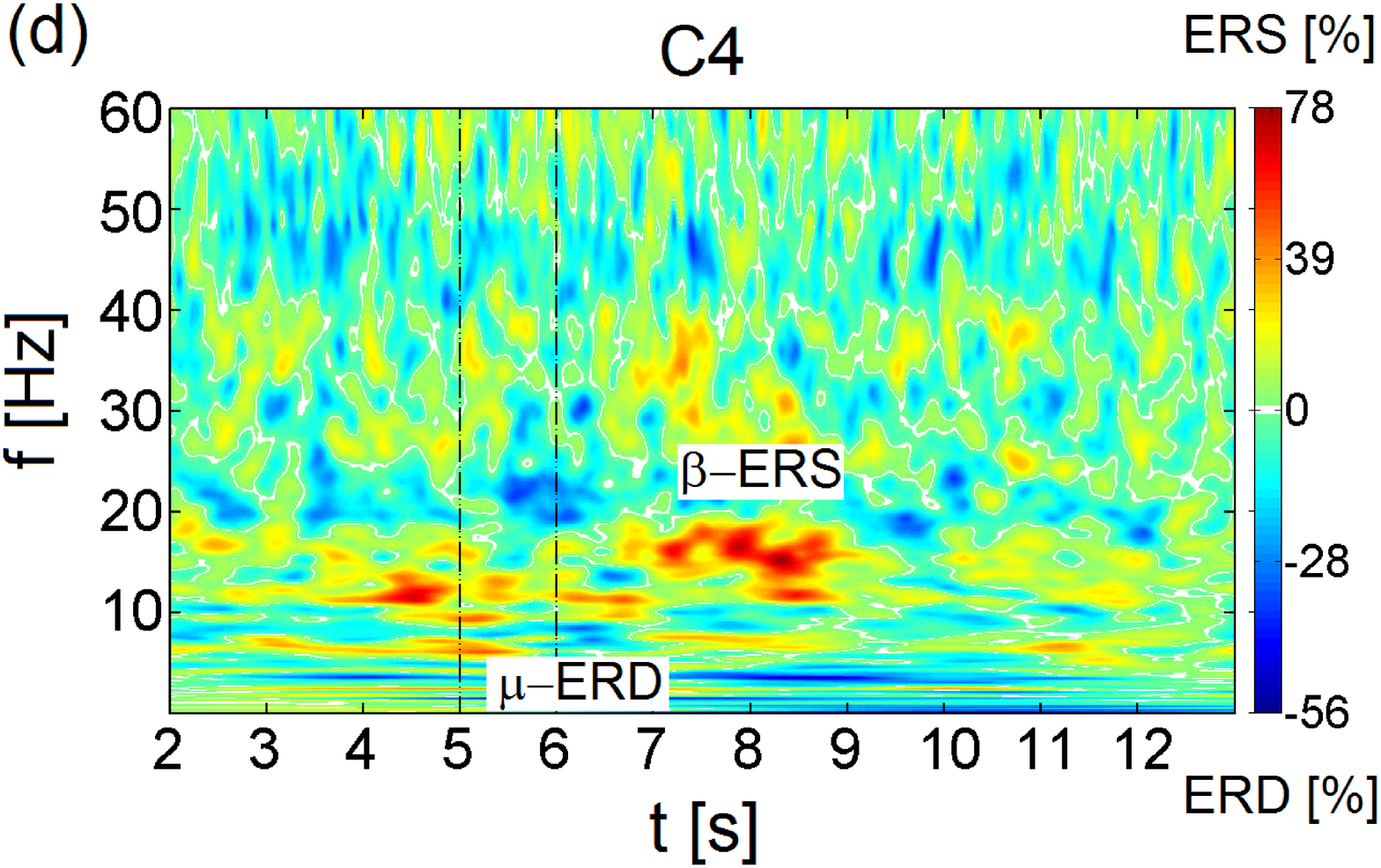}
\caption{Maps of ERD/ERS in the time-frequency plane related to the
right hand movement imagination for electrode C3 (a) and electrode C4
(b) and the left hand movement imagination for electrode C3 (c) and
electrode C4 (d). On the right side of the graph the scale of ERD/ERS
changes expressed in percent. The time interval of sound signal, which
indicates the imagination of hand movement, is marked by vertical dashed
lines. The estimation of the time-frequency distribution of the energy
density has been scalogram. The reference period is 1-2 s.}
\label{fig3}
\end{center}
\end{figure}
The EEG signal recorded during the experiment has been prepared for
ERD/ERS and FNS analysis in several steps. First, all trials with
evident artefacts have been excluded. Second, the signal has been
temporally filtered. Then, the signal has been spatially filtered using
the small laplacian filter and finally, we have averaged the signal over
all trials. 

\subsection{Time-frequency maps of ERD/ERS related to the hand movement imagination}

The ERD/ERS maps of the signal recorded during the imagination of the
right hand movement are presented in the Fig.~\ref{fig3}(a)
and~\ref{fig3}(b), at electrodes C3 and C4 respectively. Analogous, the
ERD/ERS maps of the signal recorded at electrodes C3 and C4 during the
imagination of the left hand movement are presented in the
Fig.~\ref{fig3}(c) and~\ref{fig3}(d). The time interval of the sound
signal which indicates the imagination of hand movement is marked by
vertical dashed lines. In Fig.~\ref{fig3}(a) and~\ref{fig3}(b) it can be
observed that the $\mu-$ERD, before and during the imagination, appears
bilaterally, but mainly at the contralateral electrod C3.
Simultaneously, in the range $20-30$~Hz the contralateral
synchronization ($\beta-$rebound) appears after the end of the task from
8 to 10 second. This phenomenon results from the synchronization of the
neuron's beta-activity immidiatelly after the termination of the task
performance at the contralateral side of the brain. The ERD/ERS maps
related to the imagination of the left hand movement [see
Fig.~\ref{fig3}(c) and~\ref{fig3}(d)] show that the $\beta-$rebound
appears at electrode C4 (18~Hz) as well as the electrode C3 (about
20~Hz). Desynchronization in the $\mu$ range at the contralateral
electrode is difficult to characterize and is only slightly outlined in
the vicinity of 10~Hz. On the other hand at electrod C3
desynchronization in this range is distinctly evident.
\begin{figure}[ht]
\begin{center}
\includegraphics[scale=.2]{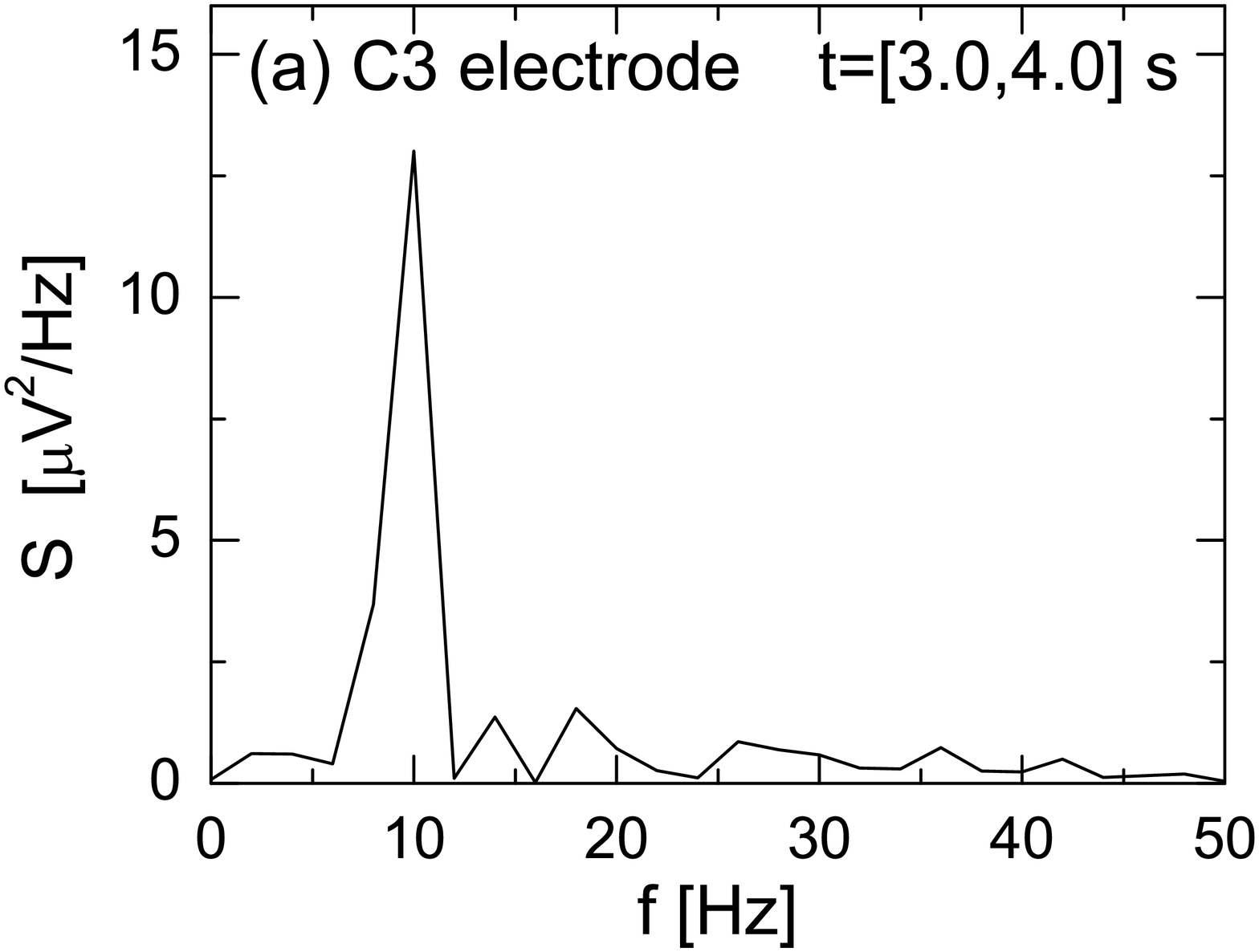}
\includegraphics[scale=.2]{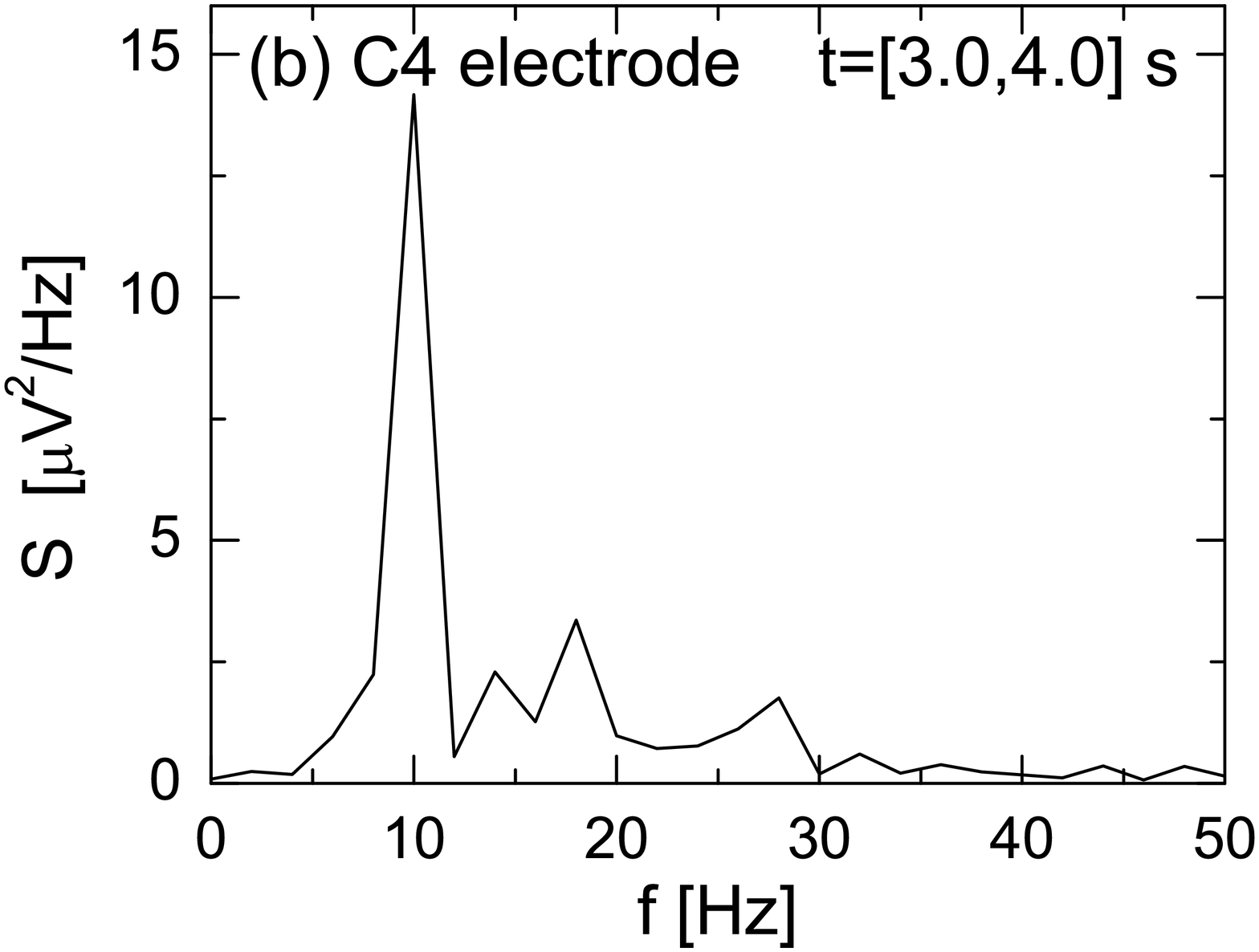}
\includegraphics[scale=.2]{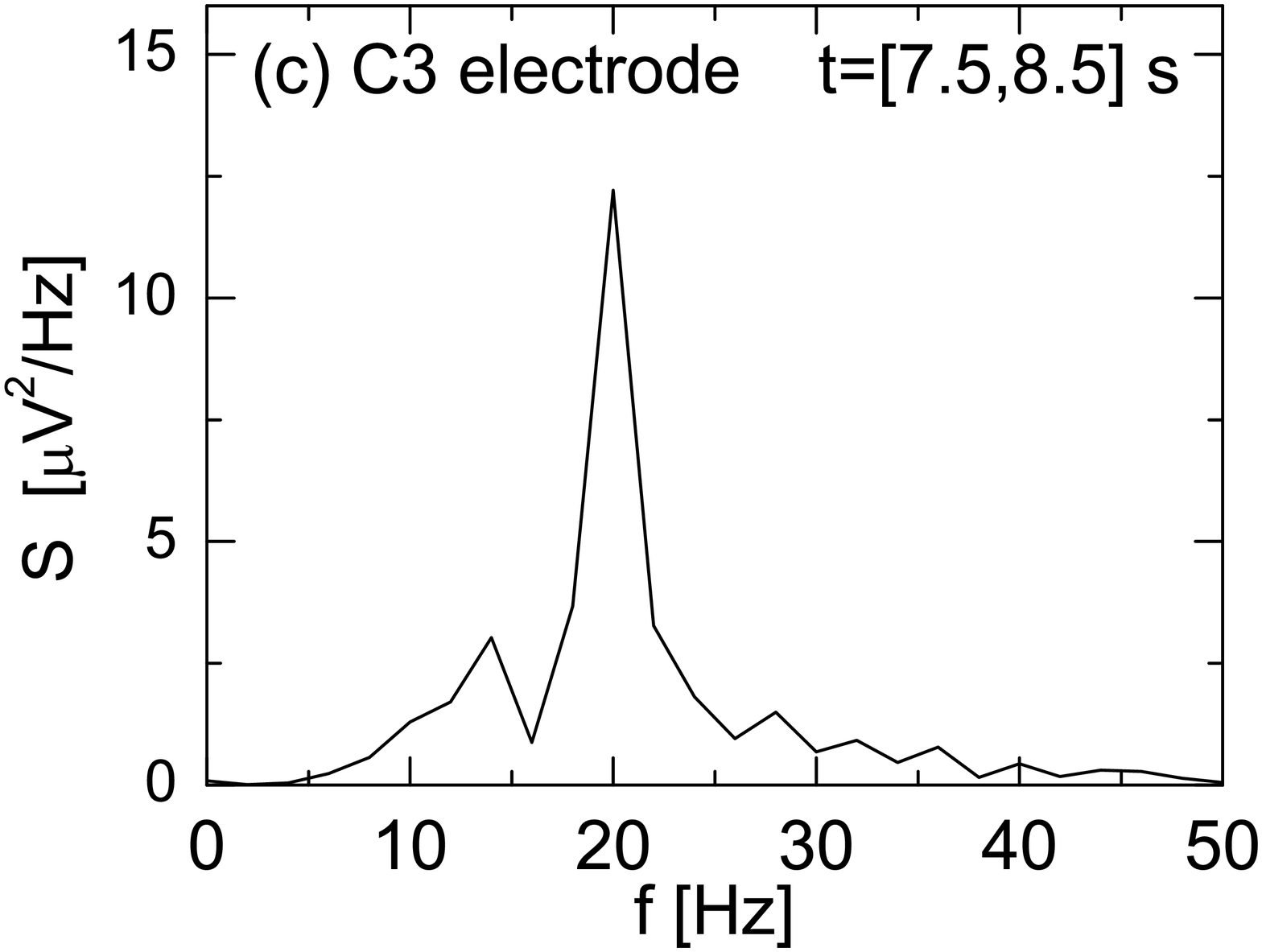}
\includegraphics[scale=.2]{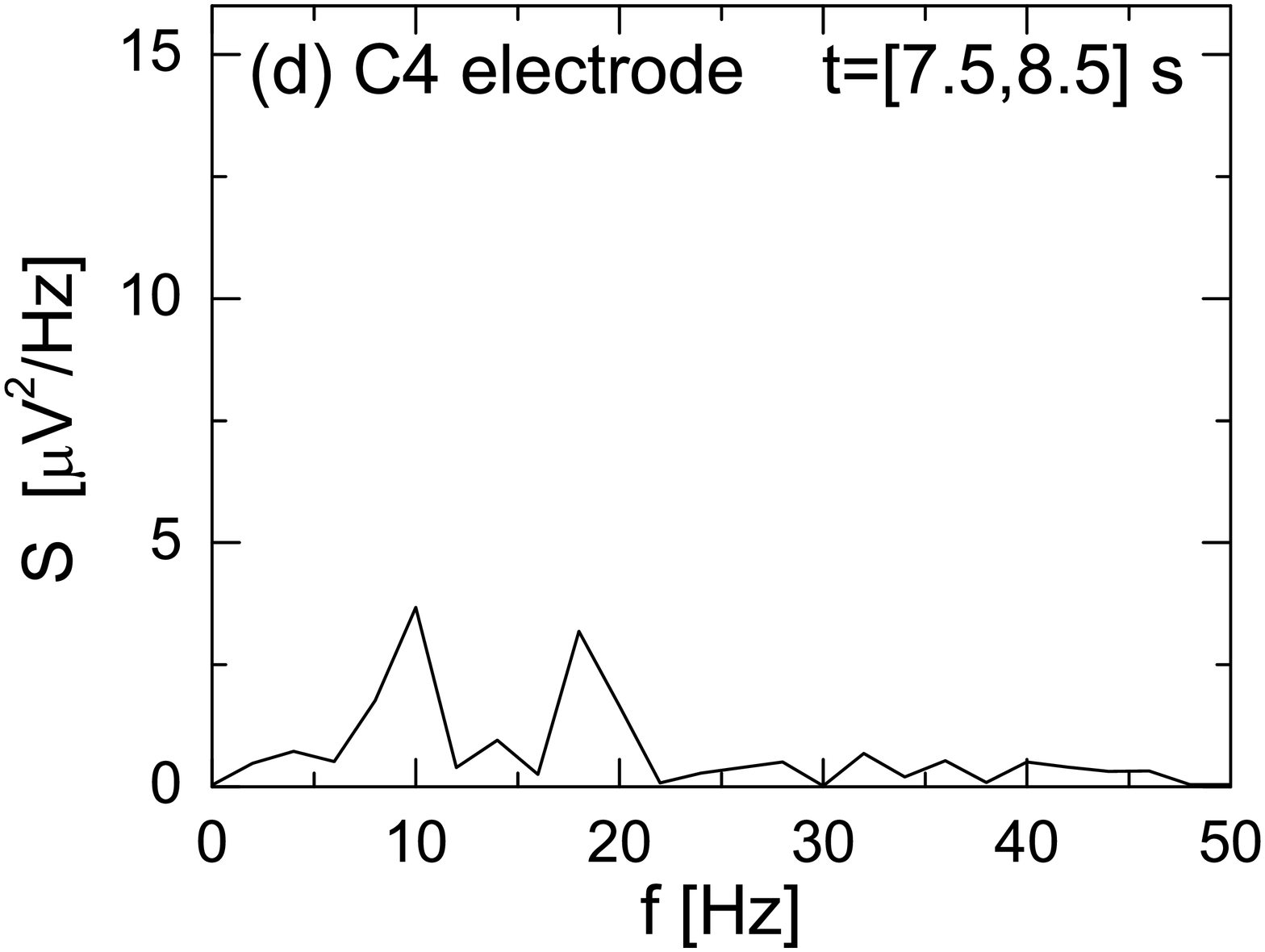}
\caption{Linear-scale power spectrum $S(f)$ of EEG signal (the cosine
transform of the autocorrelation function) calculated for two time
intervals chosen during the experiment with the right hand movement
imagination. Graphs (a) and (b) for the time interval 3$-$4~s for
electrodes C3 and C4 respectively, graphs (c) and (d) for interval
7.5$-$8.5~s for electrodes C3 and C4.}
\label{fig4}
\end{center}
\end{figure}
These results indicate that the changes in the synchronization of
rhythms occurrs contralaterally only for the imagination of the movement
with dominat hand i.e. right for the considered subject. For the
imagination of the movement with non-dominat hand i.e. left, changes at
the ipsilateral side (C3 electrode) have similar character as at the
contralateral side (electrode C4). Therefore, the imagination of the
movement with left hand, non-dominat for this patient, causes relatively
similar changes at both electrodes. Their lateralization is not evident.

\subsection{FNS parameterization}

In order to determine the parameters according to FNS methodology, the
power spectrum $S(f)$ of EEG signals has been calculated. 
\begin{figure}[ht]
\begin{center}
\includegraphics[scale=.2]{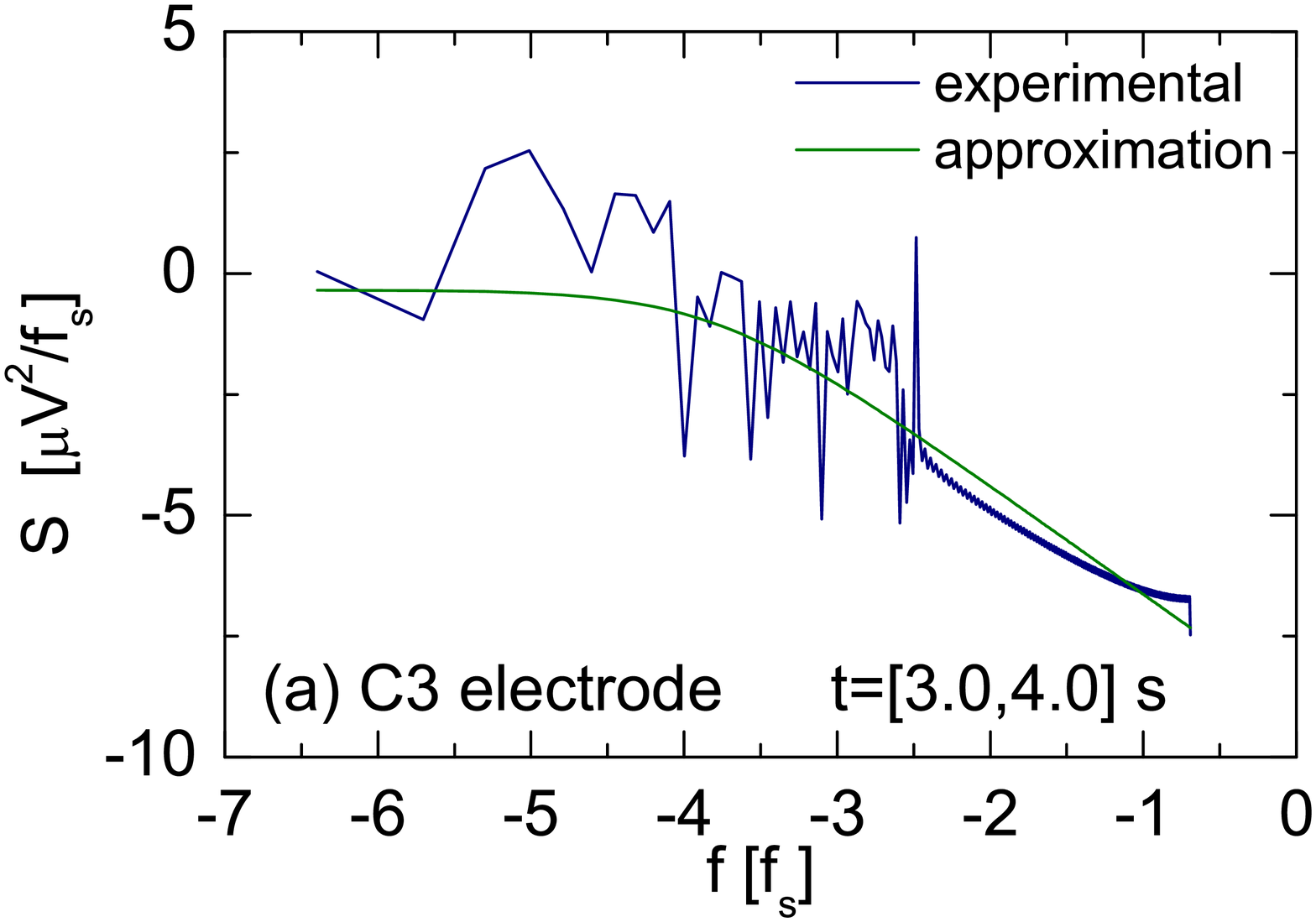}
\includegraphics[scale=.2]{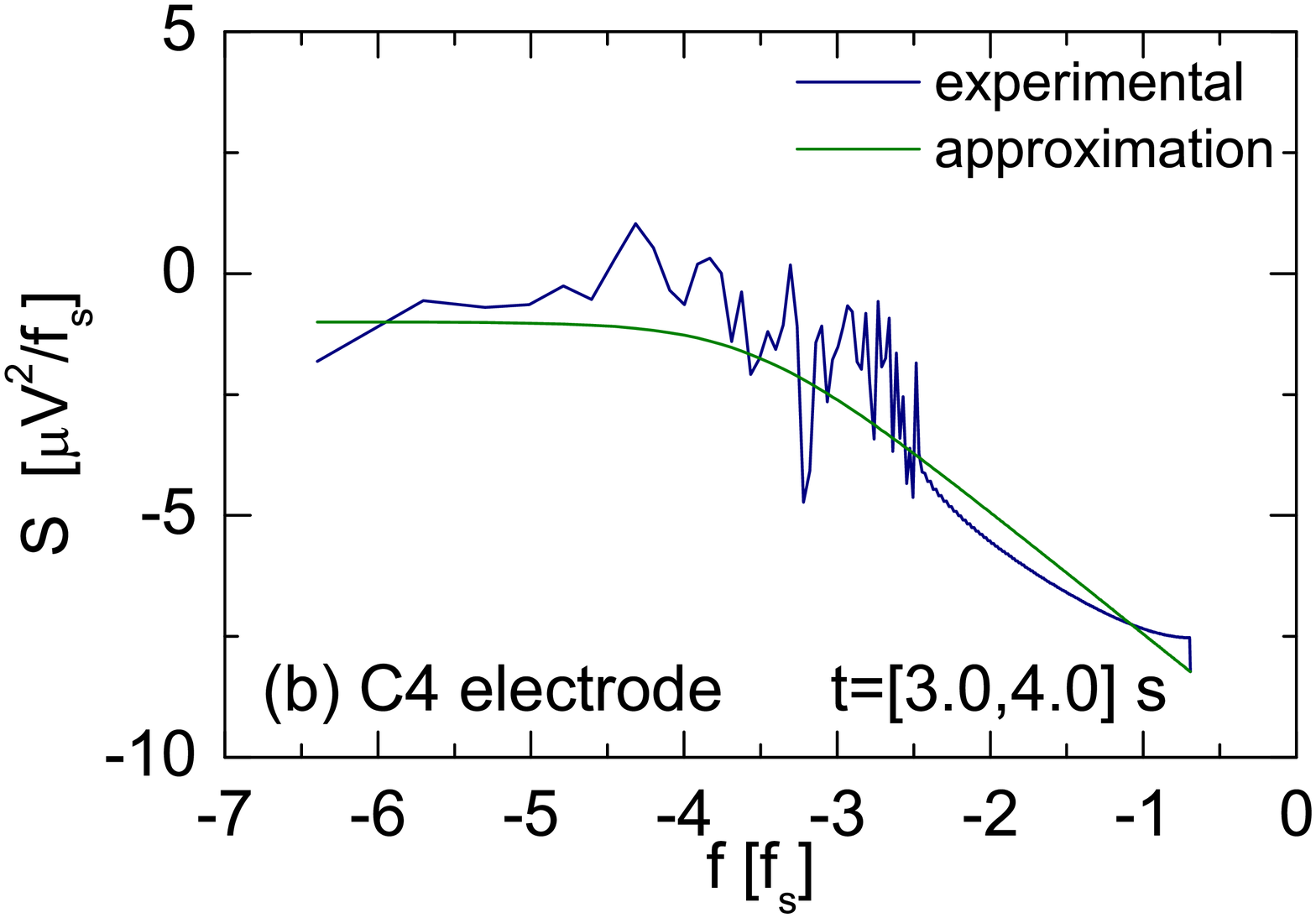}
\includegraphics[scale=.2]{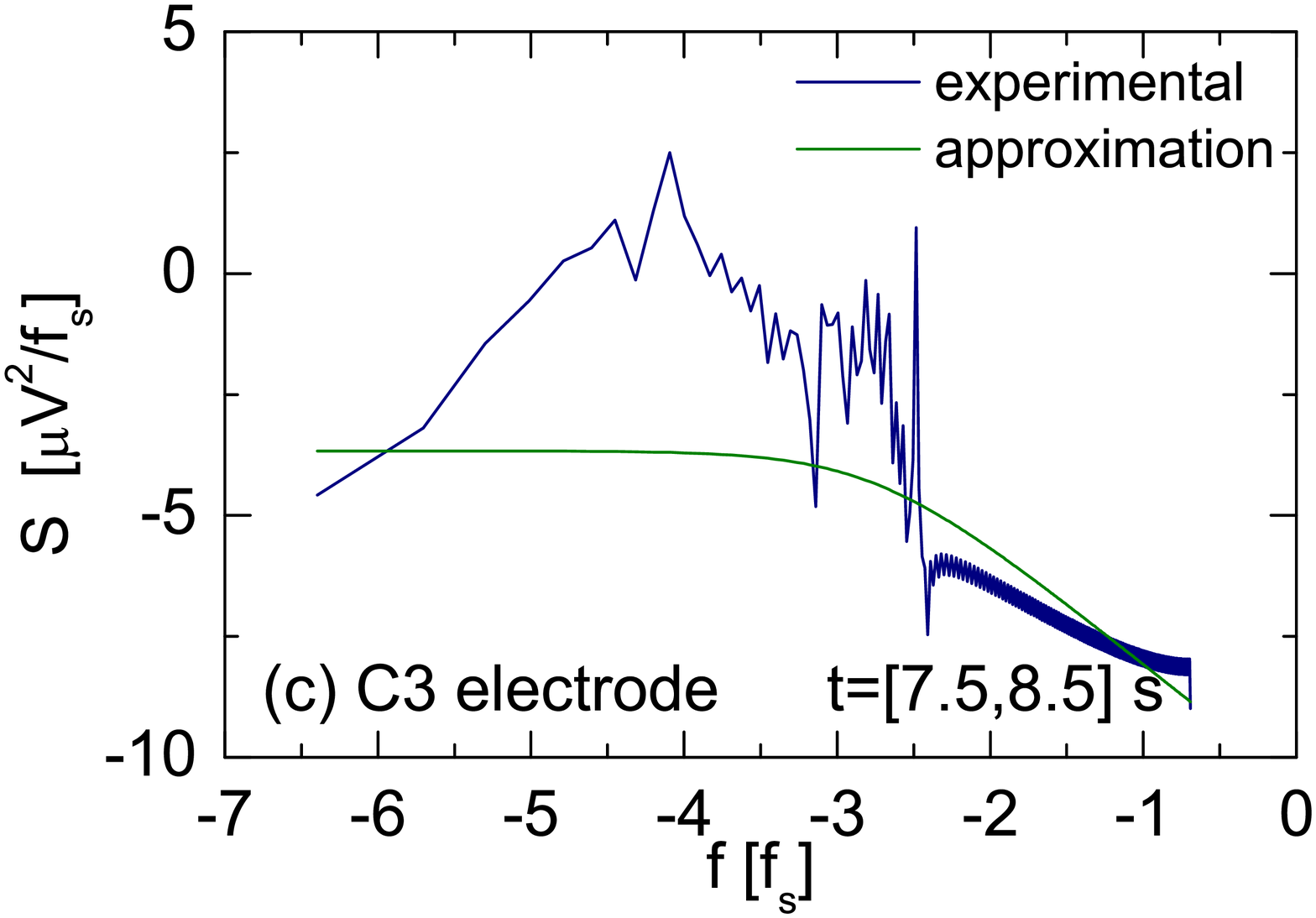}
\includegraphics[scale=.2]{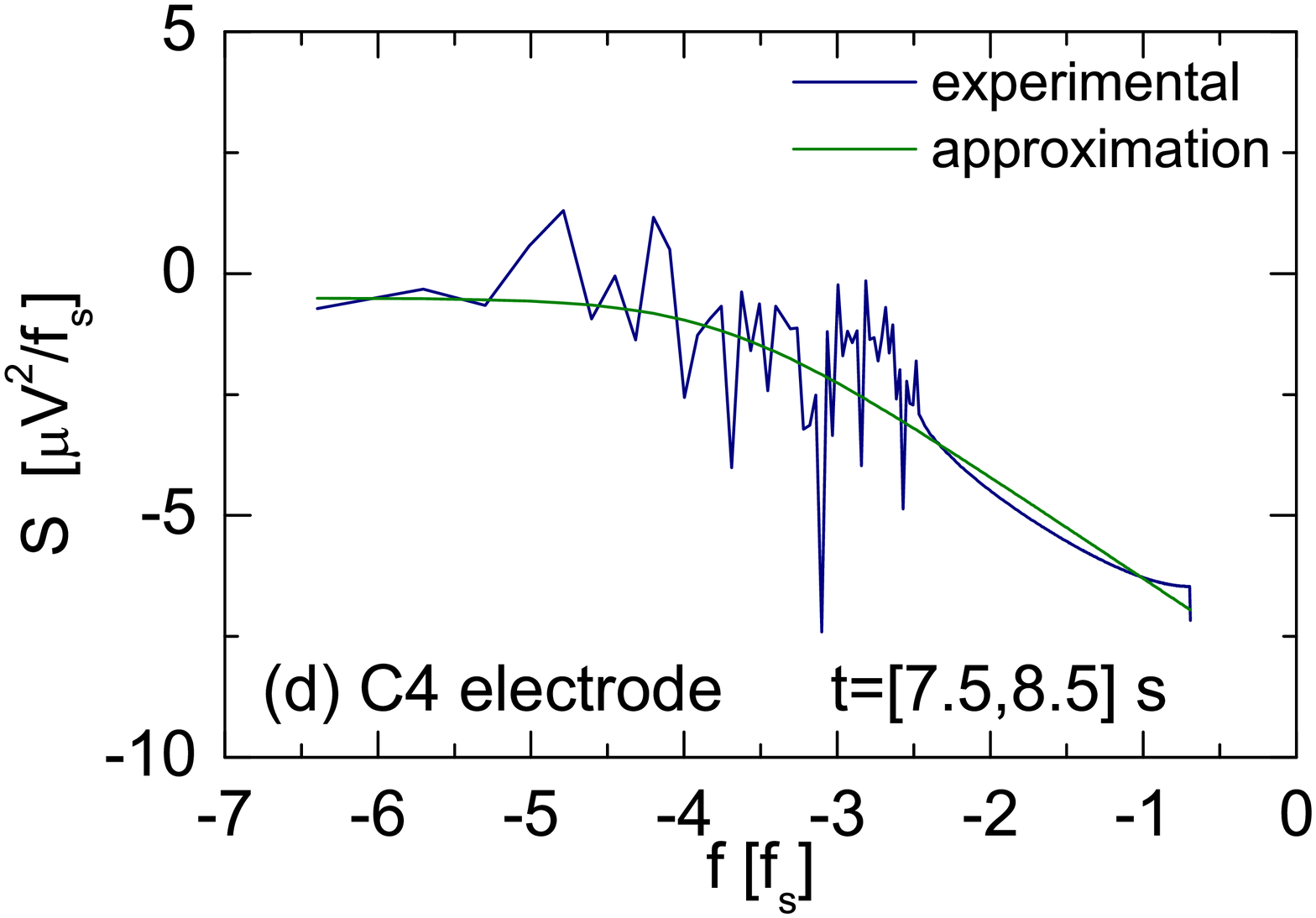}
\caption{Log-log-scale power spectra $S(f)$ of EEG signal determined for
the same data as in Fig.~\ref{fig4}. In blue $-$ the power spectra for
the experimental data, in green color $-$ the approximation of the
chaotic component calculated using the function $S_{cS}(f)$ [see
Eq.~(\ref{eq:8})].}
\label{fig5}
\end{center}
\end{figure}
In Fig.~\ref{fig4} the power spectra (the cosine transform of the
autocorrelation function) calculated using Eq.~(\ref{eq:2}), for two
different time intervals are shown. Figures~\ref{fig4}(a) and
~\ref{fig4}(b) present the power spectra of the signal recorded in a
time interval from 3 to 4 seconds at C3 and C4 electrodes, respectively.
Since the stimulus initiates the movement imagination at $5^{th}$
second, the chosen time interval is related to the preparation to the
task execution. At both electrodes the decreasing character of the EEG
signal spectrum as a function of the frequency is visible with the
distinct $\mu$ peak in 10~Hz. The peak corresponds to the dominance of
the $\mu$ wave before the task execution. The $\mu$ rhythm is then
reduced with intention to move and can be observed as ERD at the
time-frequency maps (see Fig.~\ref{fig3}).
In Figs.~\ref{fig4}(c) and~\ref{fig4}(d) the power spectrum of the
signal recorded from 7.5 to 8.5 seconds at C3 and C4 electrodes is
presented. In this time interval the strong dominance of the $\beta$
rhythm at the C3 electrode in 20~Hz can be observed. This corresponds to
the contralateral synchronization ($\beta-$rebound) visible in the
time-frequency maps after the end of the task (see Fig.~\ref{fig3}).
\begin{figure}[ht]
\begin{center}
\includegraphics[scale=.2]{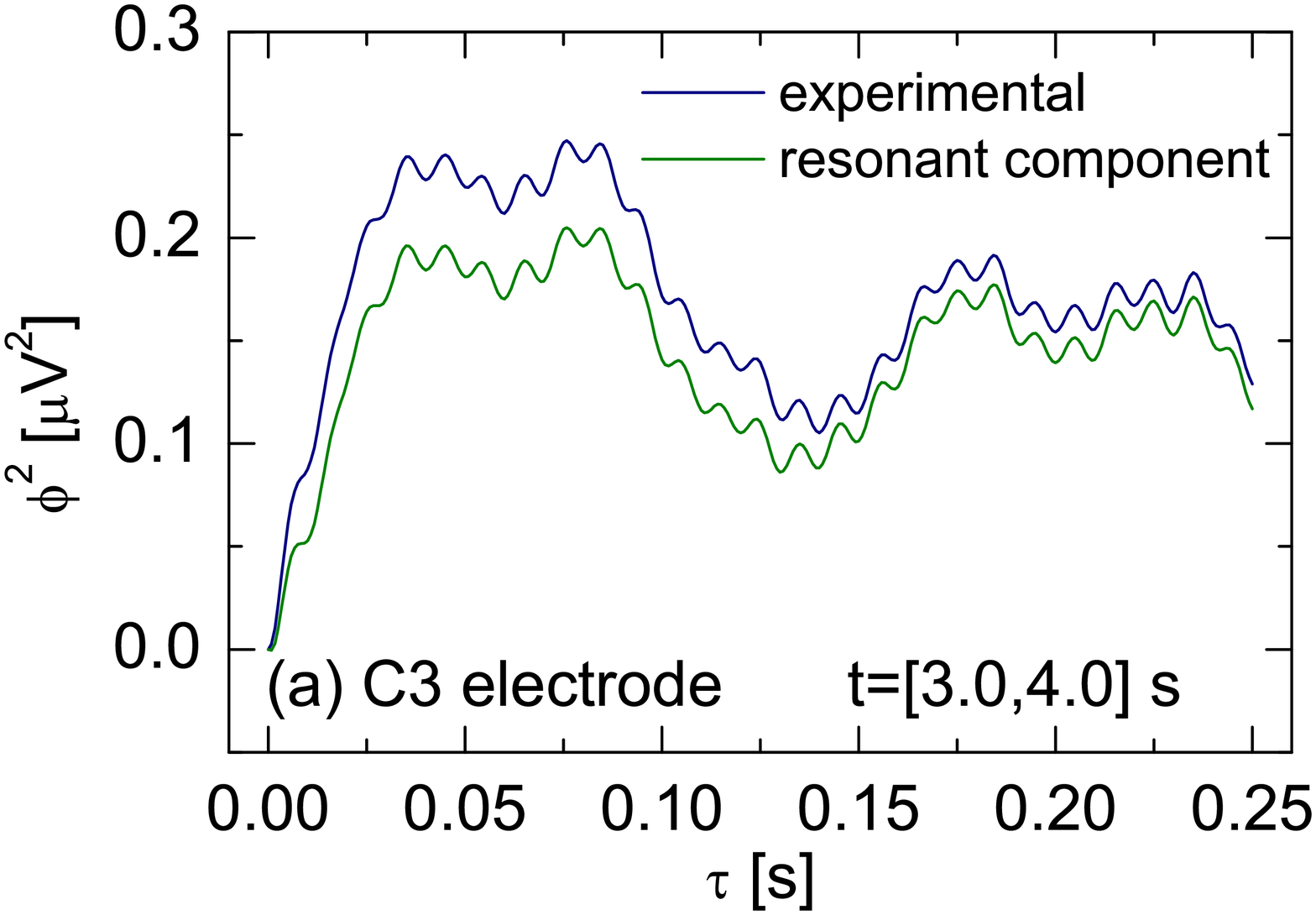}
\includegraphics[scale=.2]{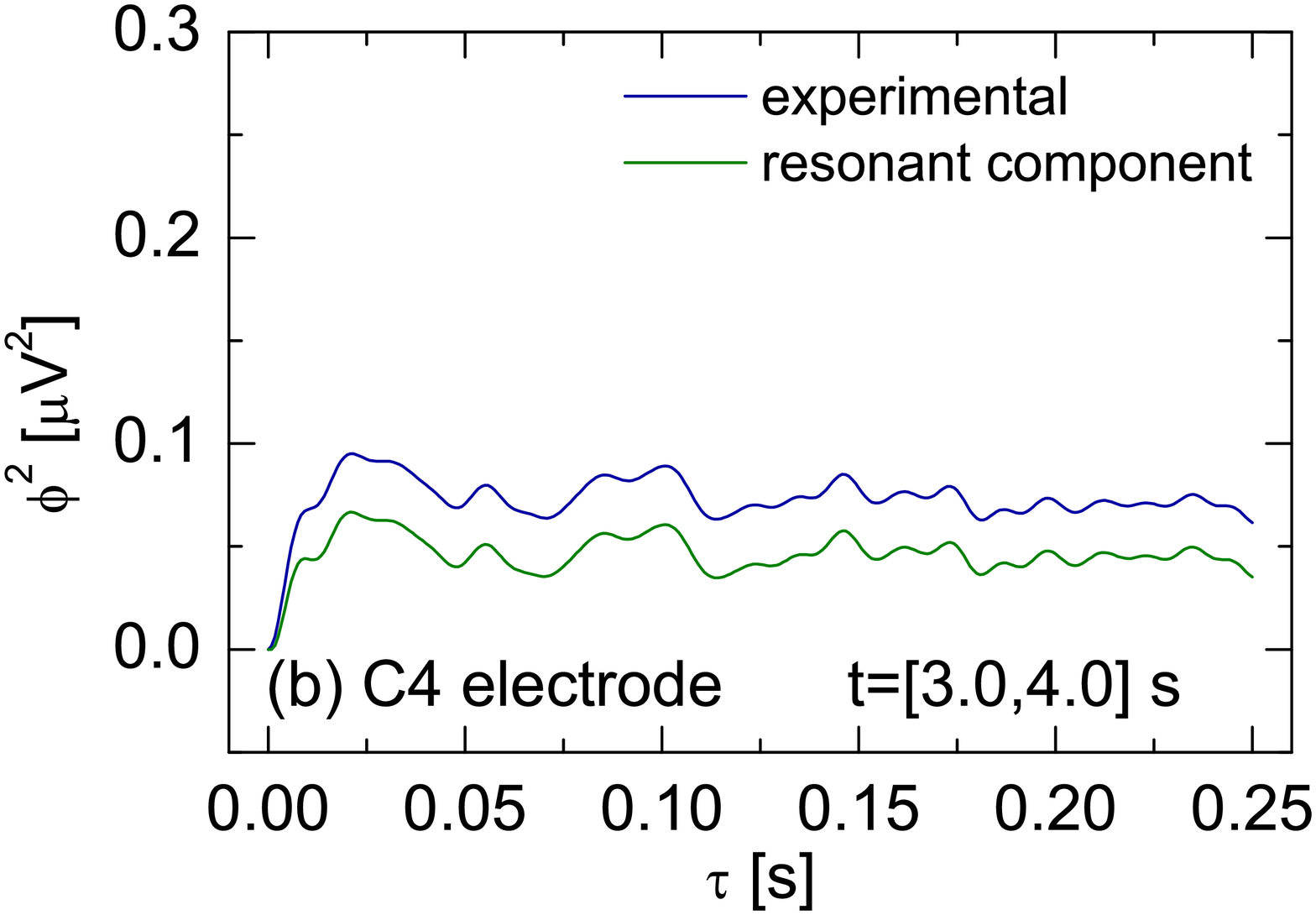}
\includegraphics[scale=.2]{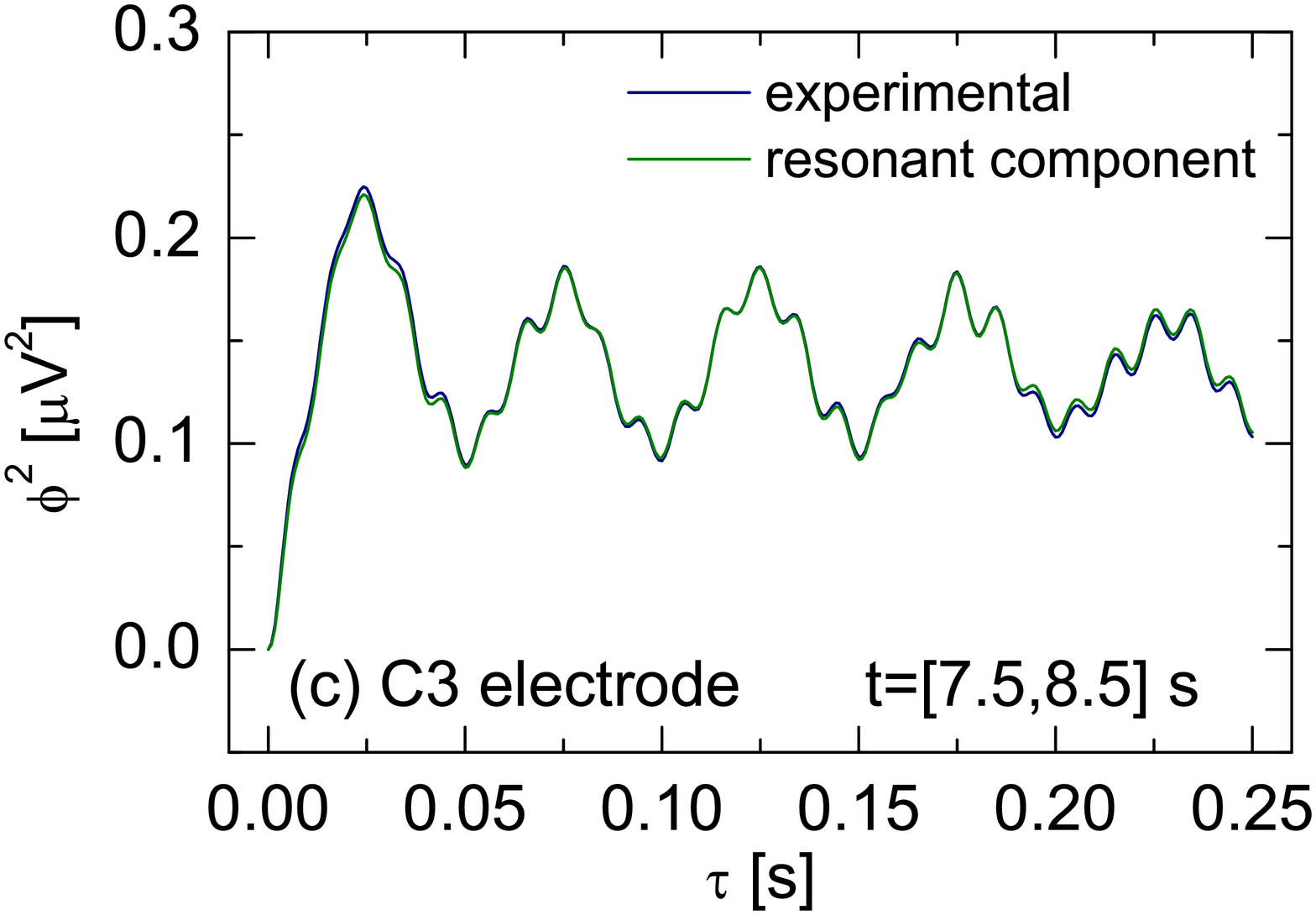}
\includegraphics[scale=.2]{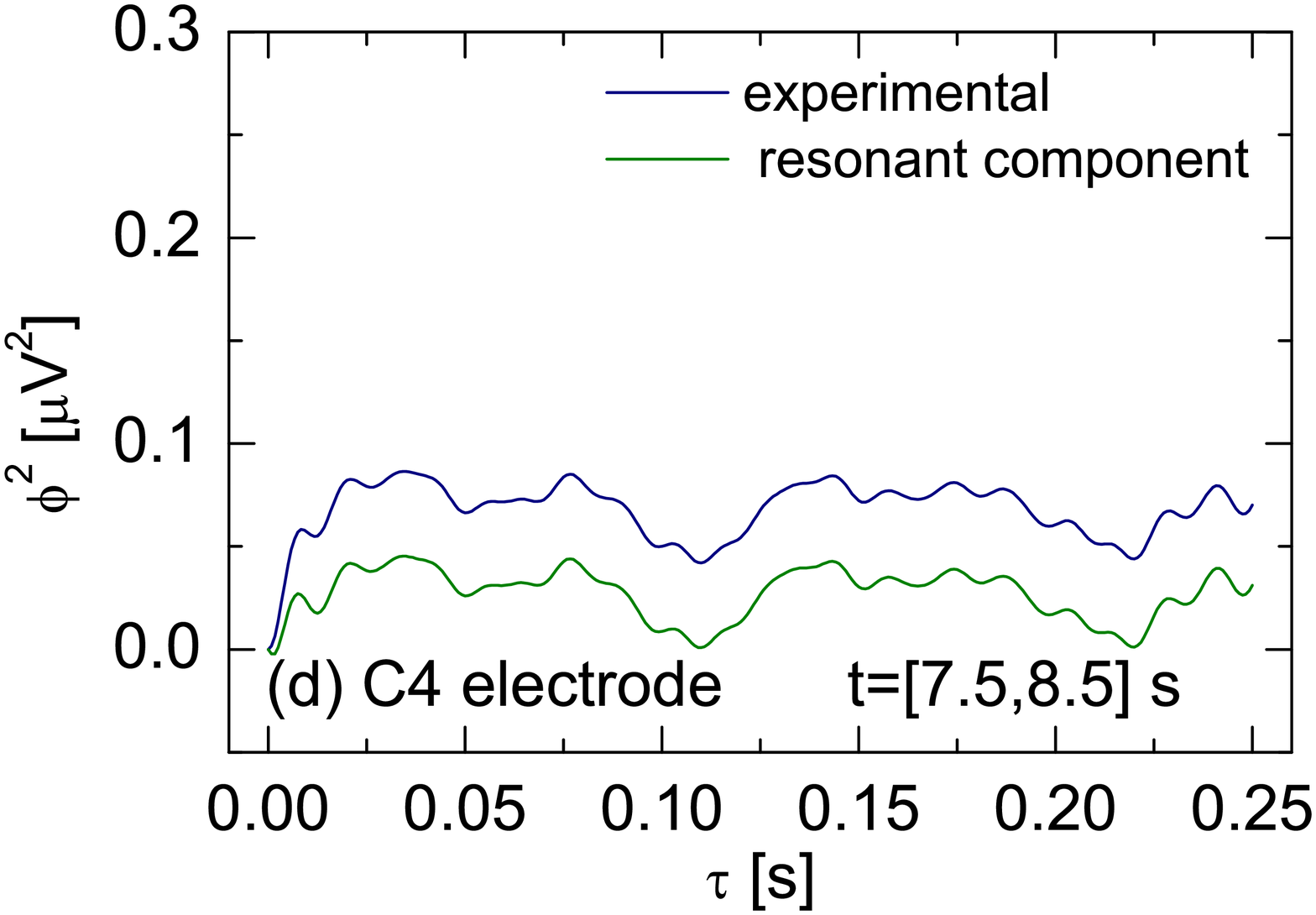}
\caption{The difference moments $\phi^2(\tau)$. Diagrams 6 (a)$-$(d) are
presented for the same data as in Fig.~\ref{fig4}. The function
$\phi^2(\tau)$ calculated for experimental data is presented by the blue
line while the green line displays the resonant component
$\phi_r^2(\tau)$.}
\label{fig6}
\end{center}
\end{figure}
\begin{figure}[ht]
\begin{center}
\includegraphics[scale=.2]{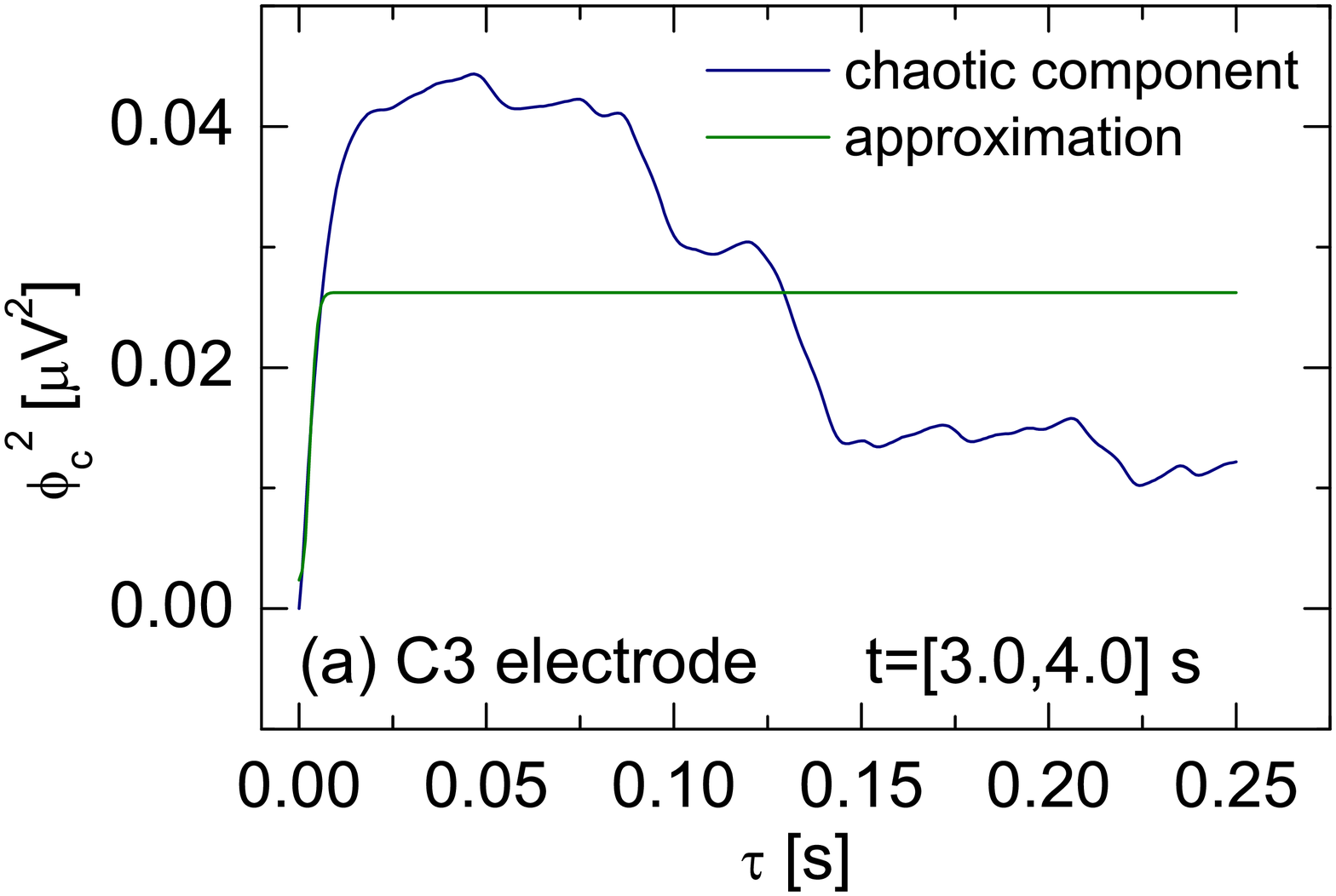}
\includegraphics[scale=.2]{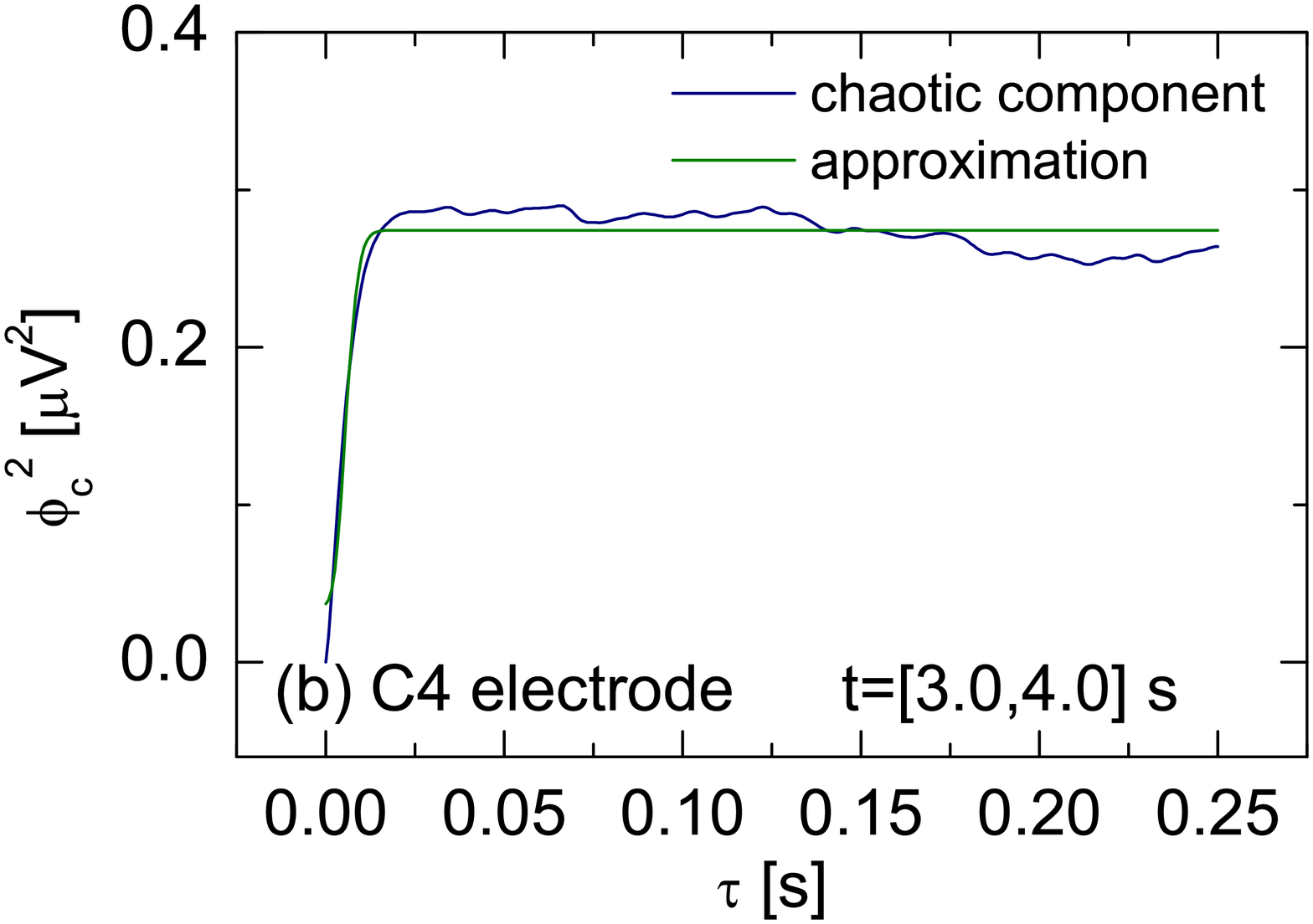}
\includegraphics[scale=.2]{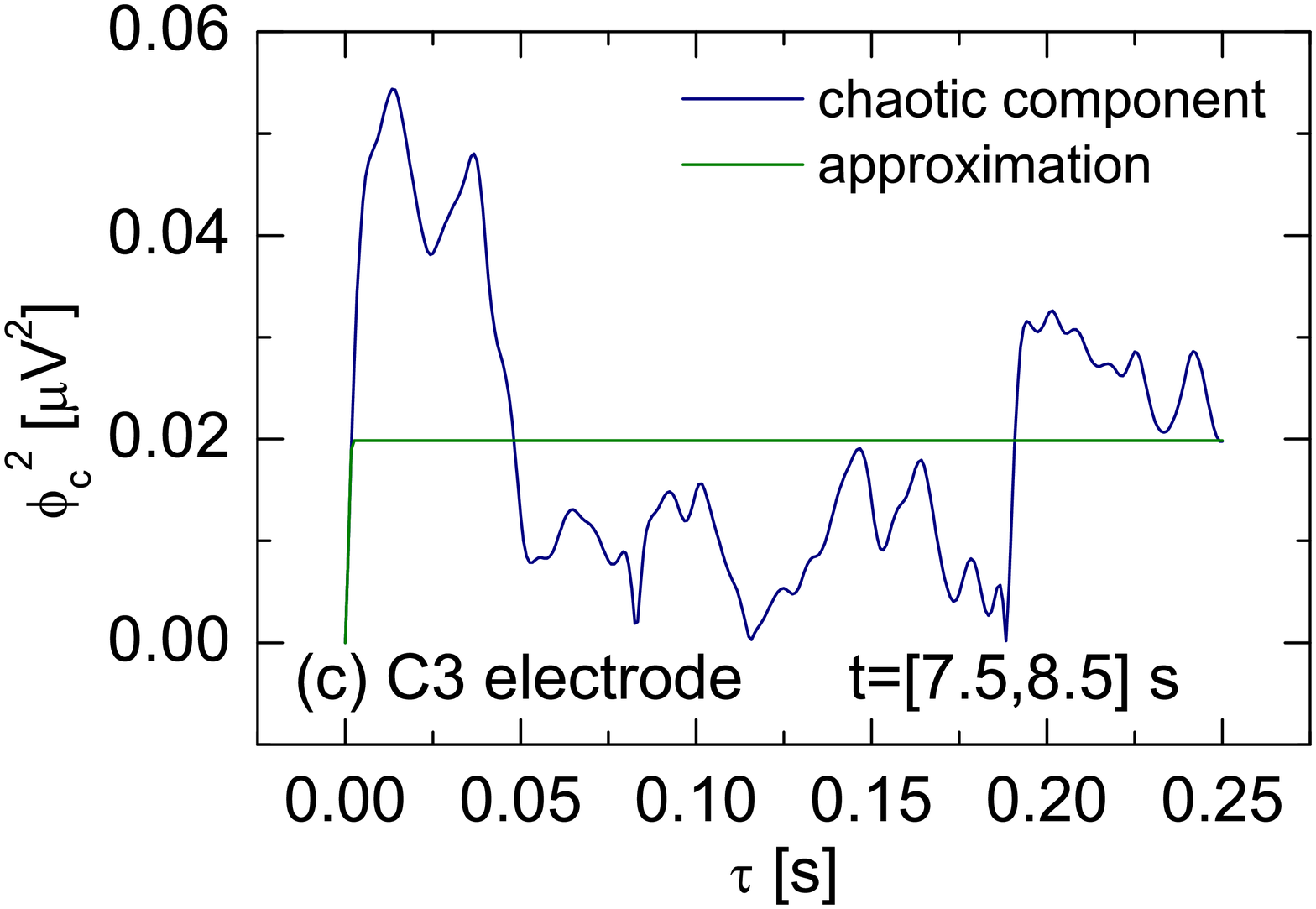}
\includegraphics[scale=.2]{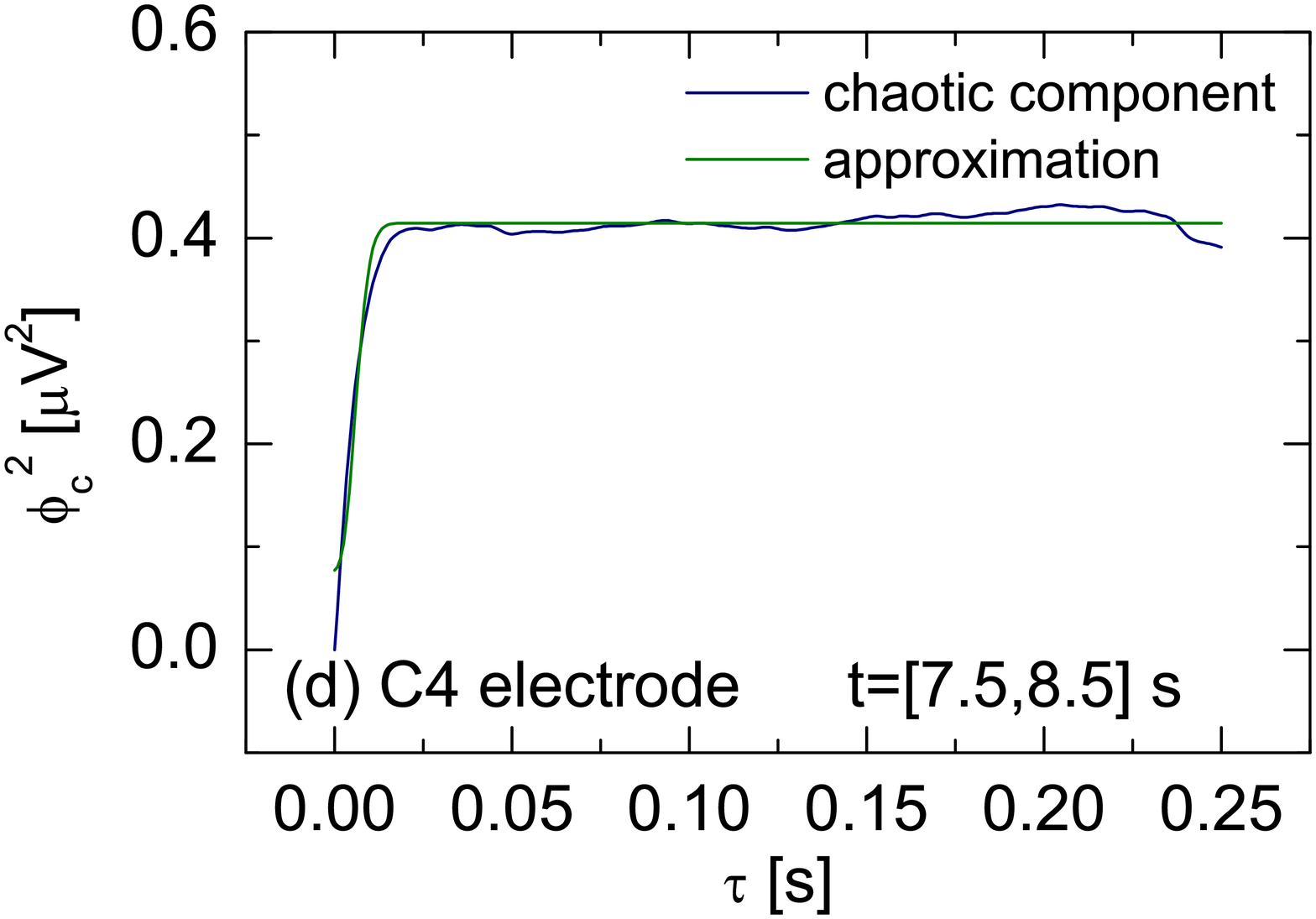}
\caption{Chaotic component of the difference moments $\phi_c^2(\tau)$
(blue line) and its approximation (green line) determined using the
formula~(\ref{eq:7}). Results are presented for the same data as in
Fig.~\ref{fig4}.}
\label{fig7}
\end{center}
\end{figure}
The FNS parameters are determined by the approximation of the chaotic
power spectrum components using the
formulas~(\ref{eq:8})$-$(\ref{eq:10}). For this purpose the power
spectra $S(f)$ of EEG signal are presented in the
logarithmic-logarithmic scale (Fig.~\ref{fig5}).
Diagrams~\ref{fig5}(a)$-$(d) are presented for the same data as shown in
Fig.~\ref{fig4}. With blue color the power spectrum of the experimental
data is shown. In the same plot, the approximation of the chaotic
component $S_{cS}(f)$ [Eq.~(\ref{eq:8})] is displayed by the green
color. The approximation has beed performed by the least squares method.
\begin{figure}[ht]
\begin{center}
\includegraphics[scale=.4]{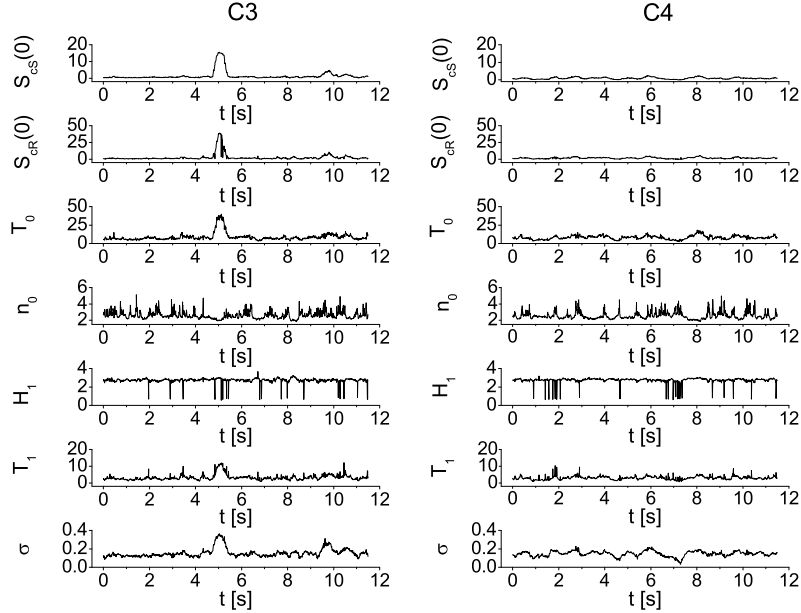}
\caption{Values of the parameters $S_{cS}(0)$, $S_{cR}(0)$, $T_0$,
$n_0$, $H_1$, $T_1$, $\sigma$ as a function of time for the imagination
of the right hand movement. Parameters have been calculated for the time
window $T=0.5$~s ($N=600$) moved along the whole duration of the
experiment. The left column shows the values of parameters for electrode
C3, right for electrode C4.}
\label{fig8}
\end{center}
\end{figure}
This procedure allows to determine $S_{cS}(0)$, $T_ 0$ and $n_0$
parameters.
The parameters $H_1$, $T_1$ and $\sigma$ are determined based on the
difference moment $\phi^2(\tau)$ presented in Fig.~\ref{fig6}. The
function $\phi^2(\tau)$ calculated for the experimental data using the
formula~(\ref{eq:5}) is displayed by the blue line. With green color the
resonant component $\phi_r^2(\tau)$ of the difference moment is marked.
The function  $\phi_r^2(\tau)$ has been determined from the resonance
component of the autocorrelation function $\psi_{r}(\tau)$ [see
Eq.~(\ref{eq:6})] calculated from the resonance component $S_r(f)$ by
applying the inverse Fourier transformation, where $S_r(f)=S(f) -
S_{cS}(f)$.
In the next step we have determined the chaotic component of the
difference moments $\phi_c^2(\tau)$ by the subtraction of the functions
$\phi_r^2(\tau)$ and $\phi^2(\tau)$. In Fig.~\ref{fig7}, the function
$\phi_c^2(\tau)$ (blue line) and its approximation (green line)
determined according to formula~(\ref{eq:7}) are presented. As
previously, the approximation has beed carried out with the least
squares method. This procedure allows to determine parameters $H_1$,
$T_1$, $\sigma$ and indirectly, using the formula (\ref{eq:10}), the
parameter $S_{cR}(0)$.

\subsection{Identification of right/left hand movement imagination}

In order to determine the changes of the FNS parameters in time, their
values have been considered in the constant time window $T=0.5$ second
(what is equivalent to $N=600$ samples), moved along the whole duration
of the experiment.
For each window position, the parameters $S_{cS}(0)$,
$S_{cR}(0)$, $T_0$, $n_0$, $H_1$, $T_1$, $\sigma$ have been calculated
with the procedure described above. Figure~\ref{fig8} displays the
values of the FNS parameters as a function of time during the
imagination of the right hand movement. Left column presents the
parameters calculated for the data recorded at C3 electrode, whereas in
the right column at electrode C4. We can observe that the values of
parameters vary with time. Figure~\ref{fig8} shows that for the data
recorded at electrode C3 an abrupt changes (visible as a peak) of the
parameters $S_{cS}(0)$, $S_{cR}(0)$, $T_0$, $\sigma$ appear at about
$5^{th}$ second, what corresponds to the initial moment of the movement
imagination. In contrary, at electrode C4, no 
meaningful changes of the parameters are observed at this moment. Since
changes related to the right hand movement imagination should
predominantly occur at electrode C3, which lies over the left hemisphere
above the area of the motor cortex, this findings agree with our
expectations.

\begin{figure}[ht]
\begin{center}
\includegraphics[scale=.4]{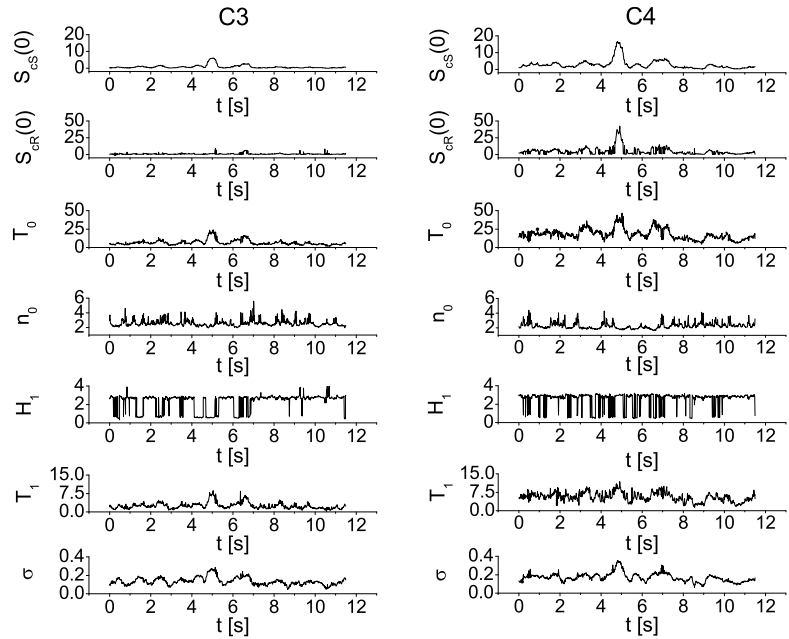}
\caption{Values of the parameters $S_{cS}(0)$, $S_{cR}(0)$, $T_0$,
$n_0$, $H_1$, $T_1$, $\sigma$ as a function of time for the imagination
of the left hand movement. Parameters have been calculated for the time
window $T=0.5$~s ($N=600$) moved along the whole duration of the
experiment. The left column shows values of parameters for electrode C3,
right for electrode C4.}
\label{fig9}
\end{center}
\end{figure}

In Fig.~\ref{fig9}, the changes of the FNS parameters as a function of
time is shown during the imagination of the left hand movement.
Significant increase of the parameter values $S_{cS}(0)$, $S_{cR}(0)$,
$T_0$, $T_1$, $\sigma$ at about $5^{th}$ second is observed at C4
electrode which lies over the motor areas contralateral to the left
hand. At the ipsilateral electrode C3, the values of parameters
$S_{cS}(0)$, $T_0$, $T_1$, $\sigma$ also increase at this moment, but
this increase is lower comparing to C4 electrode. This phenomenon is
also observed in ERD/ERS maps [see Fig.~\ref{fig3}]. It seems to be
characteristic, that for the subjects with strong right-handedness,
changes related to the movement imagination of the dominant hand occur
mainly at the contralateral electrode C3. At the same time, the
non-dominant left hand movement imagination induces changes at both
electrodes: contralateral C4 and ipsilateral C3.
\begin{figure}[ht]
\begin{center}
\includegraphics[scale=.5]{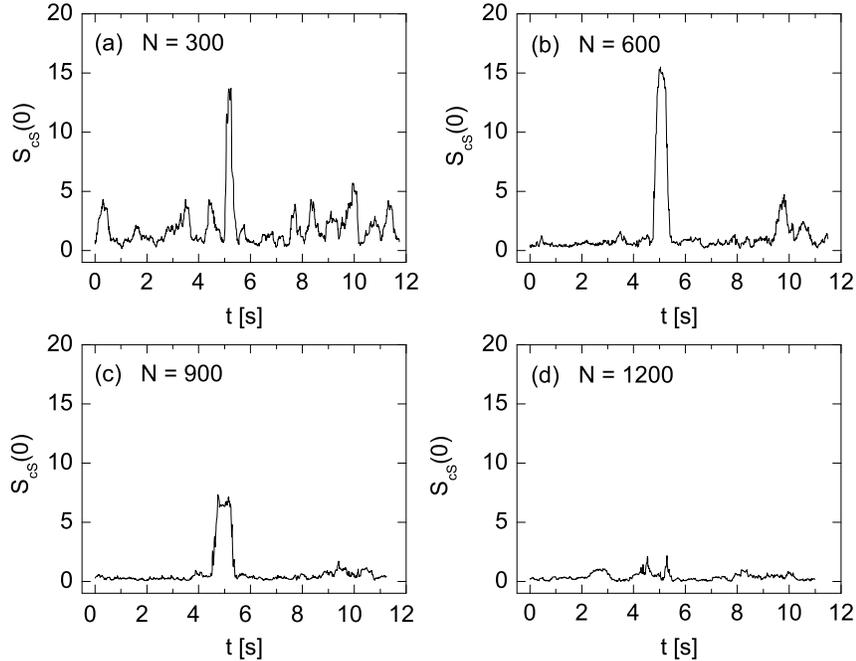}
\caption{Values of the parameter $S_{cS}(0)$ as a function of time for
imagination of right hand movement (electrode C3). Parameter has been
calculated for four different time windows i.e. (a)~$T=0.25$~s
($N=300$), (b)~$T=0.5$~s ($N=600$), (c)~$T=0.75$~s ($N=900$) and
(d)~$T=1$~s ($N=1200$).}
\label{fig10}
\end{center}
\end{figure}

Based on these results we can conclude that the investigation of the
FNS parameters as a function of time reveals the moment of the task
execution and allows to distinguish between imagination of right and
left hands movement. Nevertheless, it has to be mentioned, that the
length of the time window $T$ influences on the results. As the main
tool to extract the information contained in the signal is the
autocorrelation function, it causes that the information is averaged
over the whole given time window $T$. This means that the longer the
time window is the more information is lost as an effect of averaging.
In order to show this effect, in Fig.~\ref{fig10} the parameter
$S_{cS}(0)$ as a function of time for imagination of right hand movement
(at electrode C3) is shown. Calculations have been performed for four
different time windows i.e. (a)~$T=0.25$~s ($N=300$), (b)~$T=0.5$~s
($N=600$), (c)~$T=0.75$~s ($N=900$) and (d)~$T=1$~s ($N=1200$). We see,
that the extension of the time window $T$ above $N=1200$ 
points (what corresponds to the $T=1$~s) causes a decay of $S_{cS}(0)$
peak and loss of the information about the movement imagination.

\section{Conclusions}
\label{sec:5}

In summary, Flicker Noise Spectroscopy has been used to the analysis of
EEG signal related to the movement imagination. Three subjects performed
experiments consisted of several repetitions of either left or right
hand motor imagery. The time-frequency analysis of EEG signal from
scalp-mounted electrodes at locations C3 and C4 reveals the
event-related desynchronization (ERD) and the post-movement
event-related synchronization (ERS). The analysis of the data obtained
for all subjects have yielded the consistent results with the exception
of the individual characteristics. The signal has been parameterized in
accordance with FNS method. Parameterization of the signal with FNS
method faultless indicates the moment of movement imagination as well as
the hemisphere activated by the task. This allows to effective
differentiation between right and left hand movement imagination what is
crucial for the potential application in the brain-computer interface.

\section*{References}
%\bibliographystyle{unsrt.bst}
%\bibliography{refs}

\end{document}